\documentclass[preprint,11pt,tightenlines,showpacs,superscriptaddress,longbibliography]{revtex4-1}
\usepackage{multirow}
\usepackage{color}
\usepackage{bm}
\usepackage{ulem}
\usepackage{graphicx}
\usepackage{epsfig,psfrag}
\usepackage{amsmath}
\usepackage{amssymb}
\usepackage{amsbsy}
\usepackage{amsthm}
\usepackage{amsfonts}
\usepackage{wasysym}
\usepackage{bbm}
\usepackage{tabularx}
\usepackage{euscript}
\usepackage{enumerate}
\usepackage{amsfonts}
\usepackage{exscale}
\usepackage{bbold}
\usepackage{float}
\usepackage{slashed}
\usepackage{subfigure}
\usepackage{comment}
\usepackage[]{hyperref}

\numberwithin{equation}{section}

\def\bea{\begin{eqnarray}}
\def\eea{\end{eqnarray}}
\def\nn{\nonumber}
\def\ba{\begin{array}}
\def\ea{\end{array}}
\def\nn{\nonumber}
\def\Tr{\text{Tr}}

\def\J{\mathcal{J}}

\def\C{\mathcal{C}}
\def\L{\mathcal{L}}
\def\E{\mathcal{E}}

\def\avg#1{\left\langle#1\right\rangle}
\def\bra#1{\left\langle#1\right|}
\def\ket#1{\left|#1\right\rangle}

\def\kc#1{\left(#1\right)}
\def\kd#1{\left[#1\right]}
\def\ke#1{\left\{#1\right\}}

\def\be{\begin{equation}}       \def\ee{\end{equation}}
\def\bea{\begin{eqnarray}}      \def\eea{\end{eqnarray}}
\def\ba{\begin{array}}
	\def\ea{\end{array}}
\def\bnum{\begin{enumerate} }
	\def\enum{\end{enumerate}}

\def\nn{\nonumber}

\def\=>{\Rightarrow}
\def\>{\rightarrow}

\def\eye2{Fathbb{I}}

\def\eff{\mathrm{eff}}

\def\Tr{\mathrm{Tr}}

\DeclareMathOperator\sech{sech}

\begin{document}

\title{Complexity growth of operators in the SYK model and in JT gravity}

\author{Shao-Kai Jian}
\email{skjian@umd.edu}
\affiliation{Condensed Matter Theory Center, Department of Physics, University of Maryland, College Park, Maryland 20742, USA}

\author{Brian Swingle}
\email{bswingle@umd.edu}
\affiliation{Department of Physics, Brandeis University, Waltham, Massachusetts 02453, USA and Condensed Matter Theory Center, Department of Physics, University of Maryland, College Park, Maryland 20742, USA}

\author{Zhuo-Yu Xian}
\email{xianzy@itp.ac.cn}
\affiliation{CAS Key Laboratory of Theoretical Physics, Institute of Theoretical Physics,
Chinese Academy of Sciences, Beijing, 100190, China}

\begin{abstract}
The concepts of operator size and computational complexity play important roles in the study of quantum chaos and holographic duality because they help characterize the structure of time-evolving Heisenberg operators. It is particularly important to understand how these microscopically defined measures of complexity are related to notions of complexity defined in terms of a dual holographic geometry, such as complexity-volume (CV) duality. Here we study partially entangled thermal states in the Sachdev-Ye-Kitaev (SYK) model and their dual description in terms of operators inserted in the interior of a black hole in Jackiw-Teitelboim (JT) gravity. We compare a microscopic definition of complexity in the SYK model known as K-complexity to calculations using CV duality in JT gravity and find that both quantities show an exponential-to-linear growth behavior. We also calculate the growth of operator size under time evolution and find connections between size and complexity. While the notion of operator size saturates at the scrambling time, our study suggests that complexity, which is well defined in both quantum systems and gravity theories, can serve as a useful measure of operator evolution at both early and late times.

\end{abstract}
\maketitle

\tableofcontents

\section{Introduction}

Depending on the timescales of interest, several interrelated concepts have recently been proposed to characterize the presence or absence of chaos in quantum dynamics. The basic expectation in a quantum chaotic system is that states in the Schr\"odinger picture and operators in the Heisenberg picture become more elaborate as time passes. For the purposes of this work, we wish to compare and contrast two ways to quantify this growth: the notion of size and the notion of complexity. We define these notions in detail below, but in brief, size is a measure of how many degrees of freedom are involved in a state or acted on by an operator while complexity refers to the number of elementary steps of some type needed to prepare a state or implement an operator. These two concepts are certainly interrelated in various ways, for example, a certain minimal complexity is required in order for an operator to have large size. In this work, we study and compare precise versions of these notions in two models: the Sachdev-Ye-Kitaev (SYK) model and 2d Jackiw-Teitelboim (JT) gravity.

It is useful to consider two regimes of time, corresponding to times before or after the system has come to approximate global equilibrium. The crossover time between these two regimes, called the scrambling time, will be defined in detail below. Roughly speaking, it refers to the time after which a small perturbation has spread over the entire system.

\begin{figure}
    \centering
    \includegraphics[width=14cm]{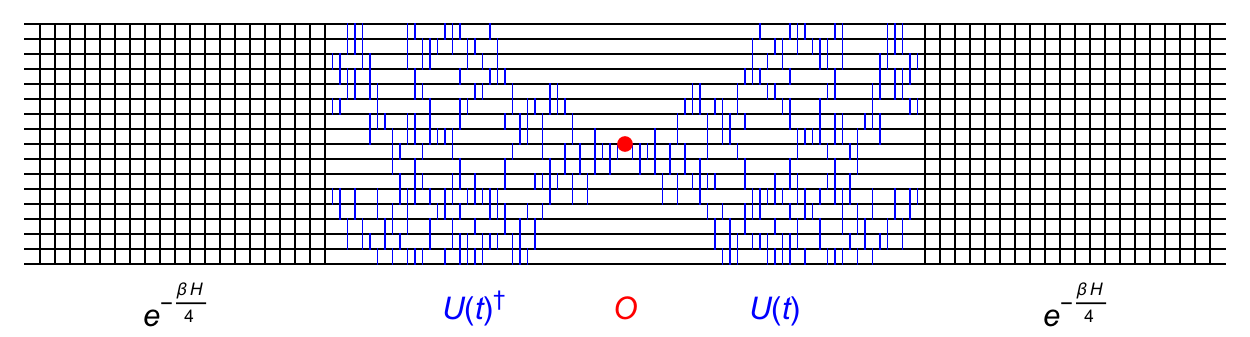}
    \caption{A cartoon of a Heisenberg operator inserted at the half of the thermal circle. Some gates of $U(t)^\dag$ and $U(t)$ on the unaffected channels have been canceled with each other, due to the switchback effect.}
    \label{fig:TN}
\end{figure}

Prior to the scrambling time, out-of-time order correlation functions (OTOCs)~\cite{Kitaev:2015a, Maldacena:2016remarks} characterize the chaotic growth of Heisenberg operators of the form
\bea\label{Heisenberg}
	O_\beta(\varphi) \equiv e^{- \kc{1 - 2\varphi/\pi}\beta H/4} O e^{- \kc{1 + 2\varphi/\pi}\beta H/4},\quad \varphi=\theta+iu,
\eea
where $H$ is the Hamiltonian, $\beta$ is the inverse temperature, $\theta$ labels the location of the insertion in the imaginary-time evolution, and $u=2\pi t/\beta$ is the real time in the unit of $\beta/2\pi$. For a simple operator $O$ which disturbs only a few degrees of freedom, time evolution causes information about this disturbance to spread over the system whenever $O$ is not conserved, $[H,O]\neq0$, a process known as information scrambling~\cite{Sekino:2008scramblers,Hayden:2007mirrors,Hosur:2015channels,Roberts:2016design}. For all-to-all chaotic Hamiltonians, information initially spreads exponentially fast with an exponent called a quantum Lyapunov exponent, until it scrambles over the whole system~\cite{Kitaev:2015a}, as illustrated in Fig.~\ref{fig:TN}. The number of degrees of freedom affected during this scrambling process is measured by the size $n$ of the Heisenberg operator $O_\beta(\varphi)$~\cite{Roberts:2018operator, Qi:2018quantum, Nahum:2017operator,vonKeyserlingk:2017operator, Carrega:2020unveiling}.

It will be convenient to translate the language of operators into the language of states using the Choi-Jamiolkowski mapping. The space of operators acting on a Hilbert space $\mathcal H$ can be mapped to a state in two copies of the Hilbert space $\mathcal H\otimes\mathcal H$ by $O\to\ket O=O\otimes\mathbb1\ket0$ where $\ket0$ is a maximally entangled state in the doubled Hilbert space. It is convenient to fix a Hamiltonian $H$ for one copy of the system (e.g. we fix $H$ for the left system, and $H^T$ for the right system which is also denoted as $H$ without confusions) and take $\ket0 = \sum_{n} \ket{E_n} \otimes \ket{E_n}$ where $\{\ket{E_n}\}$ is a basis of energy eigenstates of $H$. This mapping is appropriate at infinite temperature; it can be extended to finite temperature using the (unnormalized) thermofield double (TFD) state $\ket{\mathbb1_\beta} =  \sum_n e^{-\beta E_n/2} \ket{E_n} \otimes \ket{E_n}$ where $\mathbb1_\beta=e^{-\beta H/2}$ corresponds to $O=\mathbb1$ in~(\ref{Heisenberg}). A general operator of the form~(\ref{Heisenberg}) is mapped to a so-called partially entangled thermal state (PETS) $\ket{O_\beta(\theta+iu)}$~\cite{Goel:2018expanding}.
The evolution of the operator $ O_\beta(\varphi) = U(t)^\dag O_\beta(\theta)U(t)$ is mapped to the evolution of the state $ \ket{O_\beta(\varphi)} \equiv U(-t)\otimes U(t)\ket{O_\beta(\theta)}$, where $U(t)=e^{-iHt}$ and the two copies evolve in opposite directions in time, e.g. with Hamiltonians $H$ and $-H$. Without the insertion of $O$,  the state $U(-t)\otimes U(t)\ket{\mathbb 1_\beta}= \ket{\mathbb 1_\beta}$ is invariant under time evolution. With the insertion of non-conserved $O$, the state is no longer invariant and the dynamics can be conveniently diagnosed using correlations between the two copies.

In the context of the SYK model made from $N$ fermions $\psi^i$ obeying $\{\psi^i, \psi^j\}= \delta^{ij}$ with $q$-body interactions, the operator growth structure is well understood in the large-$q$ limit~\cite{Roberts:2018operator,Qi:2018quantum} and the conformal limit~\cite{Lensky:2020size}. Here the notion of size $n$ of an operator is the number of elementary operators---the single Majorana operators $\psi^j$---contained in that operator. The growth of size can be detected by the decay of correlations between the two systems in the PETS state, which is equivalent to a kind of OTOC. More precisely, the size is $n[O_\beta(\varphi)]=N/2 - i\sum_j\bra{O_\beta(\varphi)} \psi^j\otimes\psi^j\ket{O_\beta(\varphi)}/Z$, where the maximally entangled state $\ket 0$ is defined by $(\psi^j\otimes\mathbb1-i\mathbb1\otimes\psi^j) \ket{0}=0,\,\forall j = 1,...,N$ and the normalization factor is $Z=\avg{O_\beta(\varphi)|O_\beta(\varphi)}$.

The notion of size growth has also been explored in the context of holography. There the TFD state $\ket{\mathbb 1_\beta}$ is dual to an eternal black hole in AdS space~\cite{Maldacena:2001eternal}. For $\theta=\pm\pi/2$, acting a simple operator $O$ on the TFD state $\ket{\mathbb 1_\beta}$ corresponds to releasing a particle on the asymptotic boundary. The gravity of the black hole forces the particle to fall into the interior of the bulk and affect the near horizon region, which is the holographic bulk counterpart of the boundary growth of size~\cite{Susskind:2018fall}. Operator size has also been conjectured to be dual to the momentum of the particle~\cite{Susskind:2018fall, Brown:2018falling,Lin:2019symmetries,Susskind:2019newton,Susskind:2020momentum,Lensky:2020size}. It is therefore interesting to compare the SYK model and JT gravity model, which are closely related in the conformal limit of low temperatures~\cite{Maldacena:2016conformal,Engelsoy:2016an}. In particular, the size-momentum relation can be studied using SL(2) generators that function as both generators of spacetime transformations and measures of size~\cite{Lin:2019symmetries,Lensky:2020size}.

Now consider the situation after the scrambling time. At these longer times, the operator size has reached its equilibrium value, but operators (and states) are still evolving unitarily in the massive many-body Hilbert space. In particular, the complexity of a Heisenberg operator, a quantity borrowed from quantum information theory, is conjectured to continue to grow with time long after the scrambing time~\cite{Stanford:2014complexity,Brown:2017secondlaw}. We illustrate the evolution after scrambling using a schematic circuit diagram in Fig.~\ref{fig:TN}. The most common type of complexity considered in this context is that of circuit complexity, which is defined as the minimal number of elementary quantum gates that are required to produce a target state or operator from a reference. In a chaotic system, it is believed that the circuit complexity of an initial state will grow linearly with time until a time of order the exponential of the system entropy~\cite{Stanford:2014complexity,Cottrell:2017simple,Brown:2017secondlaw,Kim:2017qrq,Yang:2019iav}.

Various definitions of complexity for unitary operators in quantum mechanics and quantum field theories have been proposed~\cite{Jefferson:2017sdb,Roberts:2016design,Yang:2018principles}.
In this paper, we consider a different notion of complexity for operators, the K-complexity~\cite{Parker:2018a,Barbon:2019on}, defined through a Krylov basis that is uniquely determined by the evolution Hamiltonian and the reference operator in question. Using the state-operator mapping, we can define this complexity in terms of states or operators; here we focus on PETS as the reference state. The Krylov basis is obtained as follows. In the operator language, one repeatedly applies the Liouvillian map $[H,\cdot]$ to an operator $O_\beta(\theta)$ to generate a basis of operators. In dual state representation, one generates a sequence of new states starting from a PETS, e.g. $\ket{[H, O_\beta(\theta)]} = ( H \otimes  \mathbb 1 - \mathbb 1 \otimes H) \ket{O_\beta(\theta)} $. Note that, in the state language, the evolution generated by $H \otimes  \mathbb 1 - \mathbb 1 \otimes H$ is called boost evolution due to the interpretation of this transformation as a boost in AdS/CFT. It is one of the SL(2) generators mentioned above. Given this sequence of states (or operators), one then orthonomalizes them to produce the desired Krylov basis,
\bea
\ket{O_{\beta, j}(\theta)} \propto (H \otimes  \mathbb 1 - \mathbb 1 \otimes H)^j \ket{O_\beta(\theta)} + ...,
\eea
where $j$ denotes the number of applications of the Liouvillian~\footnote{The Krylov basis generated by the Liouvillian can be incomplete in the Hilbert space $\mathcal{H}\otimes \mathcal{H}$.}, and $...$ comes from the Gram-Schmidt orthonormalization process. The target state can then be decomposed into the Krylov basis $\ket{\Psi_T(t)} = \sum_j i^j \phi_j(t) \ket{O_{\beta,j}(\theta)}$, where $\phi_j$ denotes the amplitude at $j$-th basis and the factor of $i^j$ is for convenience. The K-complexity $\C_K$ is then defined by declaring that basis element $j$ has complexity $j$,
\bea\label{eq:ck-basis}
    \C_K\ket{O_{\beta,j}(\theta)}=j\ket{O_{\beta,j}(\theta)},
\eea
so the average K-complexity of the target state $\ket{\Psi_T(t)}$ is
\bea\label{eq:ck-def}
	C_K[\ket{\Psi_T(t)}]= \bra{\Psi_T(t)}\C_K\ket{\Psi_T(t)} = \sum_{j} j |\phi_j(t)|^2.
\eea

K-complexity is somewhat analogous to a circuit complexity definition in which the elementary operation is not a unitary but a Hermitian generator of unitary evolution, the Liouvillian or boost generator $H \otimes  \mathbb 1 - \mathbb 1 \otimes H$. The complexity can be defined as an operator by fixing its eigenvectors and eigenvalues in terms of~(\ref{eq:ck-basis}). Then one has a notion of average complexity by expanding general states in the Krylov basis~(\ref{eq:ck-def}). Unlike the circuit complexity, K-complexity is uniquely determined by the reference operator and the Hamiltonian, without the introduction of a set of basic gates and a tolerance parameter. This advantage is crucial when considering its gravitational correspondence, since the dictionaries of operator and Hamiltonian are clear, whereas the artificial basic gates and the tolerance parameter are not. We will give more details of the definition K-complexity in the SYK model in the next section.

It was shown in~\cite{Parker:2018a} that the K-complexity grows exponentially with time before the scrambling time for a variety of chaotic Hamiltonians. At these early times, each application of the Liouvillian will increase the operator size by at most a constant amount. As a result, the Krylov basis has a close relation with the operator size at early times. Indeed, it turns out that the K-complexity bounds any properly defined notion of operator size~\cite{Parker:2018a}. After the scrambling time, the authors of~\cite{Barbon:2019on} conjectured that the K-complexity of a chaotic system will continue to grow, now linearly with time, until it reaches a value that is exponential in the system size. The argument in~\cite{Barbon:2019on} is based on the eigenstate thermalization hypothesis (ETH) applied to the chaotic Hamiltonian of interest~\cite{Deutsch:1991quantum, Srednicki:1994chaos, Rigol:2008thermalization}. Here we explicitly show that the late-time growth of the K-complexity is bounded by a linear function. We also numerically evaluate the K-complexity growth in the SYK model and indeed find linear growth after the scrambling time. Combined with previous results, our results demonstrate that K-complexity shows a exponential-to-linear growth pattern in the SYK model. We conjecture that this holds true for all chaotic systems, namely that after a short ``dissipation time'' typically set by the local energy scales of the problem, K-complexity shows a universal exponential growth dictated by a Lyapunov exponent which then gives way to a linear growth after the scrambling time. Although we do not see a saturation of the K-complexity at late time due to the limited working precision in the calculation, the late time saturation of K-complexity is seen in a recent preprint~\cite{Rabinovici:2020operator} after we post our results.

To complete the background for our story, we consider the role of complexity in holography. In that context, circuit complexity has received enormous recent attentions for its potential relation to various features of the dual holographic geometry, especially wormholes inside black holes~\cite{Susskind:2014rva}. Two conjectures on the duality of the complexity of a state were proposed. The complexity-volume (CV) conjecture states that the complexity is proportional to the volume of the maximal spatial surface (meaning spacetime codimension-one manifold) connecting the boundary of the dual eternal black hole~\cite{Stanford:2014complexity}. The complexity-action (CA) conjecture states that the complexity is proportional to the action of the Wheeler-DeWitt patch~\cite{Brown:2015action,Brown:2015equalaction}. The proposal of holographic complexity inspires new understanding on the complexity in field theories \cite{Fu:2018nonlocal,Yang:2017nfn,Yang:2019udi} and in gravity \cite{Carmi:2016comments,Brown:2018JT,Carmi:2017time,Cai:2017sjv,Cai:2016xho,An:2018Gauss-Bonnet,Cai:2020revisit,Yang:2019gce,Yang:2019alh}.

In models with partial holographic duals, such as the SYK model, it is interesting to compare microscopic notions of complexity to holographic proposals. For the situation of interest to us, namely PETS states in the SYK model, we consider eternal black holes in JT gravity with a matter field which is dual to the operator $O$ defining the PETS~\cite{Goel:2018expanding}. In this model of gravity, the problem of gravitational backreaction of the matter field can be mapped to the motion of particles in a hyperbolic space~\cite{Maldacena:2017diving,Goel:2018expanding}. Focusing on the CV duality for simplicity~\cite{Stanford:2014complexity,Brown:2018JT}, the holographic complexity becomes proportional to the geodesic distance between the two boundaries of the eternal black hole in JT gravity. Interestingly, this holographic complexity under time evolution also exhibits an exponential-to-linear growth behavior for PETS. Similar behaviors of the complexity appear in the shock wave geometry, which caused by the matter falling into the black hole \cite{Susskind:2019newton,Susskind:2020momentum}.

Because of the similarities between the microscopic K-complexity in the SYK model and the ``coarse grained'' complexity defined through the CV conjecture, it is very interesting to compare these two notions of complexity in detail. At early times, both complexities exhibit an exponential growth at a rate set by a quantum Lyapunov exponent. In JT gravity, what appears is the maximal Lyapunov exponent $2\pi/\beta$, with $\beta$ is the inverse temperature. In the SYK model, one finds a temperature dependent Lyapunov exponent in the large-$q$ approximation, and in the conformal limit, this exponent approaches the same $2\pi/\beta$ exponent as in the gravity. We find that the K-complexity precisely matches the holographic complexity up to an unimportant constant before the scrambling time. Similarly, both complexities grow linearly after the scrambling time but with different slopes. The slope in the SYK model is set by a microscopic energy scale while the slope in JT gravity is set by the temperature. Nevertheless, both slopes are extensive in the system size and if we consider the rate of complexity growth ratio, i.e.,
\bea
    R(t) = \frac{d \log C(t)}{d t},
\eea
where $C$ refers to the complexity, then the two notions of complexity turn out to have the same rate up to times that are exponential in system entropy.

To summarize, both size and complexity provide useful windows into quantum chaotic dynamics depending on the timescale of interest. Still, it would be convenient if there is a single quantity that could capture the relevant physics of both quantities. This is not a implausible request since as the size of an operator grows, the complexity is increasing. Indeed, it has been proposed that the growth of operator size is proportional to the rate of increase of its complexity~\cite{Susskind:2014switchback}. Actually, three seemingly distinct quantities in holography---operator size, complexity, and radial momentum---are proposed to be closely related to each other~\cite{Susskind:2020momentum,Susskind:2019newton,Brown:2018falling,Barbon:2019momentumcomplexity}, as schematically shown by the following equation,
\bea
    \frac{n}{\tilde\beta} \sim P \sim \frac{d C(t)}{d t},
\eea
where $1/\tilde \beta$ represents an energy scale that sets appropriate units, and $n$ and $P$ refer to the operator size and the momentum, respectively. To verify this relation, we also carry out a calculation of operator size of the PETS in the SYK model and the corresponding SL(2) charge in JT gravity. The results show agreement with both K-complexity and holographic complexity up to the scrambling time. In this sense, the complexity serves as a useful quantity that can capture the dynamics of a simple Heisenberg operator in chaotic systems at both early and late times. While we focused on chaotic systems here, it would be interesting to explore the notion of complexity in integrable systems as well~\cite{Parker:2018a}.

The rest of this paper is organized as follows. In Sec.~\ref{sec:ck}, we explore the dynamics of the K-complexity in the SYK model. We first review the definition of K-complexity, in which the Lanczos coefficient plays an important role. The dynamics of the K-complexity is mapped to a particle moving in one-dimensional lattice made up of the Krylov basis. Then a proof concerning the late-time linear growth of K-complexity is given using a bound on the Lanczos coefficient. We also evaluate the K-complexity in the SYK model, and show that it exhibits an exponential-to-linear growth. In Sec.~\ref{sec:cv}, we calculate the holographic complexity of the PETS in JT gravity. The insertion of a simple operator causes a perturbation to the TFD state, which can be mapped to an insertion of a particle moving in the hyperbolic space. The backreaction from the operator insertion is then easily captured at the Schwarzian limit. Then the holographic complexity is measured by the geodesic connecting two asymptotic boundaries of the $AdS_2$ spacetime. The dynamics of the microscopic K-complexity and the holographic complexity share many similarities, and in certain aspect the K-complexity is a microscopic candidate of the holographic complexity. In Sec.~\ref{sec:size}, the relation between the operator size and the complexity growth rate is considered. We calculate the size of the PETS in both the SYK model and JT gravity. In particular the size is linearly related to the SL(2) charges of $AdS_2$ spacetime. We also verify that the growth rate of both K-complexity and holographic complexity of the PETS is given by its size in the Lyapunov regime. As a result, the notion of complexity is able to characterize the dynamics of Heisenberg operators in chaotic systems at both short and long times. In Appendix~\ref{appen:Ck_SYK}, we review the finite temperature generalization of K-complexity and generalize the result for non-zero inserting angles $\theta \ne 0$. In Appendix~\ref{appen:coord}, we summarize various coordinate systems of $AdS_2$. In Appendix~\ref{appen:generating}, we obtain the generating function of size operator at generic inserting angles. In Appendix~\ref{SectionScramblingTime}, we discuss the scrambling time of OTOC from the Schwarizan dynamics.

\section{K-complexity in the SYK model} \label{sec:ck}

\subsection{Review of K-complexity}

This section reviews K-complexity associated with the Krylov basis in the SYK model. Unlike an a priori basis, the Krylov basis is uniquely determined by the evolution Hamiltonian and the initial state. As a result, the K-complexity is a natural notion capturing the intrinsic dynamics of the evolution operator, without the ambiguity of choosing an operator basis or elementary gates. We take the inner product in operator space $A$ to be defined by the inner product in $\mathcal H\otimes\mathcal H$, namely $\avg{O|O'}=\Tr[O^\dag O']$, so the norm is $|| O || = \avg{O|O}^{1/2}$.

We work at infinite temperature in this section, which means that the PETS is actually $\ket{O_{\beta = 0}(\theta)}$, where $\theta$ is now irrelevant since the thermal circle is a point with no size. Hence, for notational simplicity, we neglect the subscript $\beta$ and the angle variable $\theta$, and denote the time evolved PETS by $\ket{O_{\beta = 0}(\varphi)} = \ket{O(t)}$. It is convenient to normalize this operator such that $||O|| = 1$. The Heisenberg evolution of an operator is generated by the Liouvillian $\L = [H,\cdot]$, {\it i.e.}, $\ket{O(t)} = e^{i t\L} \ket{O}$. The Krylov basis is defined through the Liouvillian superoperator:
\bea
    && \ket{O_0} = \ket{O}, \quad b_0=0, \\
	&& \ket{O_n} = b_n^{-1} \ket{A_n}, \quad \ket{A_n} = \L \ket{O_{n-1}}  - b_{n-1} \ket{O_{n-2}},\quad  b_n = ||A_n||, \quad  n \ge 1.
\eea
The second line is merely carrying out a Gram-Schmidt procedure on the states $|A_n\rangle$ to produce the states $|O_n \rangle$. The iteration stops once the Liouvillian fails to generate a linearly independent state. The set of states generated typically span a space of dimension $K$ that is of order $K \sim \dim\mathcal H^2$, but they do not always form a complete basis, $K \le \dim \mathcal H^2$. In particular, this happens when the Hamiltonian has conserved charges. For example, the SYK Hamiltonian preserves fermion parity, so the Krylov basis spans the even (odd) fermion parity subspace if one starts with an even (odd) parity reference state.

In terms of the Krylov basis, the Liouvillian superoperator is a simple tridiagonal matrix,
\bea
	L_{mn} = \langle O_m| \L |O_n \rangle = \delta_{n,m-1} b_{n+1} + \delta_{n,m+1} b_n.
\eea
The coefficients $b_n$ are also known as Lanczos coefficients.

The Heisenberg operator corresponding to $O$ can be decomposed in the Krylov basis, i.e., $\ket{O(t)} = \sum_{n=0}^m i^n \phi_n(t) \ket{O_n}$, with $\phi_n$ real. Unitary evolution implies $\sum_{n=0}^{K} |\phi_n|^2 = 1$, so $\phi_n$ can be understood as a wavefunction for the Heisenberg operator in the Krylov basis. The effective Schr\"odinger equation obeyed by the wavefunciton is
\bea\label{Schrodinger}
	\partial_t \phi_n(t) = b_n \phi_{n-1}(t) - b_{n+1} \phi_{n+1}(t), \quad \phi_0(0) = 1.
\eea
This Schr\"odinger equation (\ref{Schrodinger}) effectively describes a quantum particle moving in one dimensional chain in which each lattice site corresponds to an element of the Krylov basis. We note that this mapping from operators to quantum particles shares similar ideas with the mapping from unitary operators to points in a complexity geometry~\cite{Brown:2017secondlaw,Bao:2018virial}.

Now, the K-complexity is defined as a linear operator $\C_K$ which is diagonal in the Krylov basis and which simply counts the basis elements~\cite{Parker:2018a},
\bea
    \C_K\ket{O_n}=n\ket{O_n}.
\eea
Thus, the average K-complexity of the Heisenberg operator $O(t)$ is the average position of the particle moving in the chain, i.e.,
\bea\label{eq:ck}
	C_K[O(t)]= \bra{O(t)}\C_K\ket{O(t)} = \sum_{n=0}^K n |\phi_n(t)|^2.
\eea

The dynamics of K-complexity is governed by the Lanczos coefficient through the Schr\"odinger equation~(\ref{Schrodinger}).
Following Ref.~\cite{Barbon:2019on}, we can build intuition by considering a continuum limit of the Schr\"odinger equation obtained by introducing a short-range cutoff $\epsilon$ with $x= \epsilon n$. Expanding~(\ref{Schrodinger}) and keeping the lowest-order term in $\epsilon$, we get a continuous version of the Schr\"odinger equation,
\bea
	\partial_t \phi(x,t) = - v(x) \partial_x \phi(x,t) - \frac12 v'(x) \phi(x,t),
\eea
where the position-dependent velocity $v(x) = 2\epsilon b_n$ captures the information from the Lanczos coefficient. Using a coordinate transformation defined by $dy = \frac{dx}{v(x)}$, the wavefunction changes to $\psi(y,t)= \sqrt{v(x)} \phi(x,t)$, and the Schr\"odinger equation becomes a solvable wave equation,
\bea
	(\partial_t + \partial_y) \psi(y,t) = 0, \quad \psi(y,t) = \psi_i(y-t),
\eea
where $\psi_i(y)$ is the initial wavefunction at $t=0$. Note that $\psi$ can be also understood as a wavefunction with normalization $1= \frac1{\epsilon} \int dy |\psi(y,t)|^2$.

The average K-complexity of $\psi$ is then
\bea
	C_K(t) = \sum_n n |\phi_n(t)|^2 = \frac1{\epsilon^2} \int dx x |\phi(x,t)|^2 =  \frac1{\epsilon^2} \int dy x v(x) |\phi(x,t)|^2 = \frac1{\epsilon^2} \int dy x(y) |\psi(y,t)|^2,
\eea
where, again, $x(y)$ is determined by the coordinator transformation $dy = \frac{dx}{v(x)}$.

Given a localized initial condition corresponding to the reference state, $|\psi_i(y)|^2 = \epsilon \delta(y)$, the average K-complexity is
\bea
	C_K(t) = \frac1{\epsilon} \int dy x(y) \delta(y-t) = \frac{x(t)}\epsilon,
\eea
which is fully determined by $x(y)$ or, equivalently, by the velocity through $\frac{dx}{v(x)} = dy $.

For Lanczos coefficients given by $b_n = \alpha n^{\delta}$, the velocity is $v(x) = 2\alpha \epsilon (x/\epsilon)^\delta $. Hence, the K-complexity grows as
\bea
	C_K(t) \sim
	\begin{cases} e^{2\alpha t}, & \quad \delta =1\\
					(2\alpha t)^{1/(1-\delta)}, & \quad \delta <1
	\end{cases}.
\eea
In particular, for $\delta = 1$, the average K-complexity grows exponentially, while for $\delta=0$, it grows linearly. Ref.~\cite{Parker:2018a} showed that the Lanczos coefficients are bounded by a linear function when $n \ll N$, where $N$ is the system size, implying that the average K-complexity grows at most exponentially up to the scrambling time, $\alpha t \sim \log N$. In the next section, we show that the Lanczos coefficients are bounded by a constant when $n \gg N$, and consequently, that the average K-complexity can grow no faster than linearly in time at late times.

\subsection{Dynamics of K-complexity in chaotic systems}

To bound the Lanczos coefficients, it is useful to consider moments of the Liouvillian superoperator,
\bea
    \mu_{2n} \equiv \bra{O_0} \L^{2n} \ket{O_0},
\eea
which are closely related to the following Green function or auto-correlation function,
\bea
    G(t) &=& \frac{\Tr[O^\dag(0) O(t)]}{\Tr[O^\dag O]}= \bra{O_0} e^{i t \L} \ket{O_0} = \sum_{n} \frac{(it)^{2n}}{(2n)!} \bra{O_0} \L^{2n} \ket{O_0} = \sum_{n} \frac{(it)^{2n}}{(2n)!} \mu_{2n}, \\
    \mu_{2n} &=& \int \frac{d\omega}{2\pi} \omega^{2n} G(\omega), \qquad G(\omega) = \int dt e^{i \omega t} G(t).
\eea
Note that the Green function is normalized such that $G(t=0) = 1$.

Knowing the moments, one can get the Lanczos coefficients using an explicit relation between the two~(e.g., see Appendix A of \cite{Parker:2018a}). The relation between Lanczos coefficient and the momemt can be obtained also by using saddle-point approximation~\cite{Avdoshkin:2019euclidean}. If the Lanczos coefficients have the smooth form $b_n = \alpha n^\delta$ as $n \rightarrow \infty$, the moment is dominated by
\bea
\label{eq:saddle_moment}    \mu_{2n} &=& e^{2n \int dt [H(t) + \delta \log(2f(t)) ]} \times \alpha^{2n} n^{2\delta n}, \\
    H(t) &=& - \frac{1+f'(t)}2 \log[1+f'(t)] - \frac{1-f'(t)}2 \log[1-f'(t)], \quad 
\eea
where $f(t), 0\le t \le 1$ is a function uniquely determined by $f(0)=f(1)=0$, $f(t)>0$, $|f(t)|<1$ and $\frac{f''(t)}{1-f'(t)^2} + \frac{\delta}{f(t)} = 0$. For $\delta=1$, $f(t) = \frac{\sin\pi t}{\pi}$ which leads to $\mu_{2n} = \left( \frac{4\alpha n}{\pi e} \right)^{2n}$~\cite{Avdoshkin:2019euclidean}. For a generic $\delta>\frac12$, $f(t)$ can be solved by inverse hypergeometric function. But in any case the prefactor of~(\ref{eq:saddle_moment}) is given by $e^{\text{constant} \times n}$, so the the moments have the asymptotic behavior
\bea\label{eq:saddle}
	\mu_{2n} \approx \alpha^{2n} e^{2\delta n \log n + o[n]}, \quad n \gg 1.
\eea
where $o[n]$ denotes terms that are at least in the same order of $n$. While this calculation gives some intuition of the relation between the Lanczos coefficient and the moment, in the following, what we are going to use is the following rigorous bound~\cite{Parker:2018a},
\bea\label{eq:relation}
	\prod_{k=1}^n b_k^2 \le \mu_{2n} \le C_n \max_{\{b_k\}} (b_k^{2n}),
\eea
where $C_n = \frac{(2n)!}{n!(n+1)!}$ is the Catalan number. 

Now we show that the Lanczos coefficients are bounded by a constant for $n \gg N$. Take the Hilbert space to consist of $N$ Majoranas (with $N$ an even integer) and consider an all-to-all $q$-body Hamiltonian $H = \sum_x h_x$ such as the SYK model. Each term in the Hamiltonian is taken to be bounded, $||h_x || \le \E $. We consider the moments and the Lanczos coefficients generated from a simple operator $O$ (for example, a single Majorana operator in the SYK model). Defining $l_x = [h_x, \cdot]$, the $n$-th power of the Liouvillian is
\bea
	\L^n \ket O = \sum_{x_1,...,x_n} l_{x_n} l_{x_{n-1}} ... l_{x_1} \ket O.
\eea
Each application of $l_{x_k}$ increases the size of the operator by at most $q$, so the largest size of $l_{x_k}... l_{x_1} \ket O$ is $k q + 1$. Here, the size of a given operator refers to the number of elementary operators, such as a single Majorana operator $\psi$ in the SYK model, contained in that operator. In order to have a nonzero term $l_{x_{k+1}} l_{x_{k}} ... l_{x_1} \ket O$, $h_{x_{k+1}}$ and $l_{x_{k}} ... l_{x_1} \ket O$ should have a nonvanishing overlap. For each nonvanishing term $l_{x_k}... l_{x_1} \ket O$, applying $\L$ will lead to at most $2 (k q +1)N^{q-1}$ nonvanishing terms. As a result, the total number of nonvanishing terms of type $\L^n \ket O$ is bounded,
\bea
	\prod_{k=1}^n 2 ((k-1) q + 1) N^{q-1} < (2 q N^{q-1})^n n!.
\eea
Importantly, the number of nonvanishing terms increases as a factorial of $n$. Moreover, each individual term $l_{x_{n}} ... l_{x_1} \ket O$ is bounded by $|| l_{x_{n}} ... l_{x_1} \ket O || \le (2\E)^{n}$. So the moment is bounded by
\bea
	\mu_{2n} = || \L^n O||^2  \le (2\E)^{2n} (2 q N^{q-1})^{2n} (n!)^2 < (4 N^{q} \E)^{2n} (n!)^2.
\eea
Thus, according to the bound between the moments and the Lanczos coefficients (\ref{eq:relation}),
\bea
    \prod_{k=1}^n b_k^2 \le \mu_{2n} < (4 N^{q} \E)^{2n} (n!)^2,
\eea
the Lanczos coefficients can grow asymptotically $1 \ll k < N/q$ at most linearly, i.e., $b_k \propto k^{\delta}$, $\delta \le 1$. This is also consistent with the saddle-point calculation~(\ref{eq:saddle}).

However, when the size of $\L^n \ket O$ is greater than $N$, which occurs when $n \ge N/q$, we can improve the bound as follows. If $l_{x_{k}} ... l_{x_1} \ket O$ has size $N$, applying $\L$ can only lead to at most $2 N^{q}$ nonvanishing terms. As a result, when $n \ge N/q$, the total number of nonvanishing terms in $l_{x_{k}} ... l_{x_1} \ket O$ is
\bea
	(2 q N^{q-1})^{N/q} (N/q)! (2 N^{q})^{n-q/N} < (N/q)! (2 N^q)^n.
\eea
In contrast to the situation when $n<N/q$, the number of nonvanishing terms now increases at most exponentially with respect to $n$. The moments are bounded by
\bea
	\mu_{2n} \le (2\E)^{2n} ((N/q)!)^2 (2 N^q)^{2n} = ((N/q)!)^2 (4 N^q \E )^{2n}.
\eea
The Lanczos coefficients are bounded by the relation
\bea
	b_{N/q+1}^2 ... b_n^2 \le \frac{\mu_{2n}}{b_1^2 ... b_{N/q}^2} = \frac{((N/q)!)^2 (4 N^q \E )^{2(N/q)}} {b_1^2 ... b_{N/q}^2} (4 N^q \E )^{2(n-N/q)},
\eea
which implies that $\delta = 0$ for $n \gg N/q$. Combining above results, we have
\bea
	b_n \le \begin{cases}
		\frac{\lambda_L}2 n, \qquad & 1 \ll n \ll N/q \\
		\frac{\lambda_C}2 N, \qquad & N/q \ll n \ll 2^N
		\end{cases},
\eea
where $\lambda_L$ is the Lyapunov exponent, and $\lambda_C$ is a constant independent of $n$. The factor $N$ in the second line is to capture the system size dependence of $b_n$ at $n \gg N/q$, such that $\lambda_C$ is independent of the system size (see the following). The Lanczos coefficients of $n$ are thus bounded by a linear function in $n$ followed by a plateau that is independent of $n$. 

Now as mentioned previously, the late-time plateau of Lanczos coefficients was first discussed in~\cite{Barbon:2019on} based on the ETH conjecture. Here, we provided an explicit proof of this plateau behavior of the Lanczos coefficients, which strengthens the results of~\cite{Barbon:2019on}. As we now review, the ETH conjecture is still useful to give an estimate of the plateau value of the Lanczos coefficients~\cite{Barbon:2019on}. We continue to work in a Hilbert space of $N$ Majorana fermions, so the total dimension is $2^{N/2}$.
Using the Lehmann representation in the energy eigenbasis $|E_a \rangle$, the Green function and moments read
\bea
    G(\omega) &=& \frac1{\Tr[O^\dag O]} \sum_{ab} 2\pi \delta(\omega - (E_a - E_b)) |O_{ab}|^2, \qquad O_{ab} \equiv \langle E_a | O | E_b \rangle \\
    \mu_{2n} &=& \int \frac{d\omega}{2\pi} \omega^{2n} G(\omega) = \frac1{\Tr[O^\dag O]} \sum_{a,b} (E_a - E_b)^{2n} |O_{ab}|^2.
\eea
According to ETH, the matrix elements $O_{ab}$ can be approximated by a random matrix to high accuracy in the thermodynamic limit, i.e., $O_{ab} = A(E_a, E_a) \delta_{ab} + A(E_a,E_b) 2^{-N/4} R_{ab}$, where $A(E_a, E_b)$ is a smooth function of energies, $R_{ab}$ denote a random matrix with zero mean and unit variance. If we assume $A(E_a, E_b) = A(0,0) F(E_a - E_b)$ is a function of the energy difference only, then we have
\bea\label{muETH}
    \mu_{2n} =  2^{-N/2} \sum_{a,b} (E_a - E_b)^{2n} \frac{|A(E_a, E_b)|^2}{\sum_c |A(E_c, E_c)|^2} = 2^{-N} \sum_{a,b} (E_a - E_b)^{2n} |F(E_a - E_b)|^2 \approx (N \E)^{2n},
\eea
where we have implicitly averaged over the random matrix $R_{ab}$. The moment $\mu_{2n}$ is dominated by the largest energy difference between two many-body energy eigenvalues at large $n$. This implies $b_n \approx N \E$ at $n \gg N$, namely, the plateau value of Lanczos coefficient is proportional to the system size $N$.

The linear-to-plateau behavior of Lanczos coefficients in turn implies that, in a chaotic system, the average K-complexity of a simple Heisenberg operator as a function of time exhibits an exponential-to-linear growth,
\bea\label{eq:complexity}
	C_K(t) \approx \begin{cases}
		e^{\lambda_L t}, \quad & t_d \ll t \ll t_* \\
		\lambda_C N t, \quad & t_* \ll  t
		\end{cases}
\eea
where $\lambda_L$ and $\lambda_C$ are constants, and $t_d = \lambda_L^{-1}$, $t_* = \lambda_L^{-1} \log N/q$ are the dissipation time and the scrambling time, respectively.

\subsection{K-complexity growth of operators in the SYK model}

\begin{figure}
\subfigure[]{
	\includegraphics[width=5.2cm]{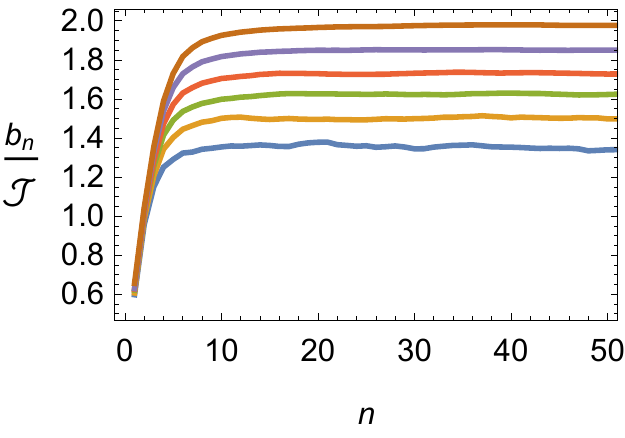}\label{SYKLanczos}} \quad \quad
\subfigure[]{
	\includegraphics[width=5.2cm]{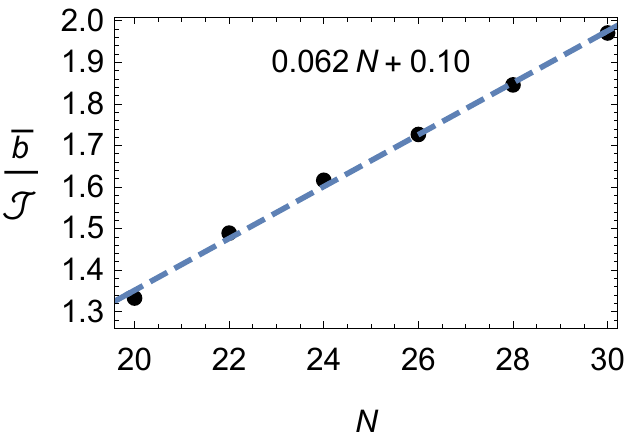}\label{fig:plateau}} \\
\subfigure[]{
	\includegraphics[width=5cm]{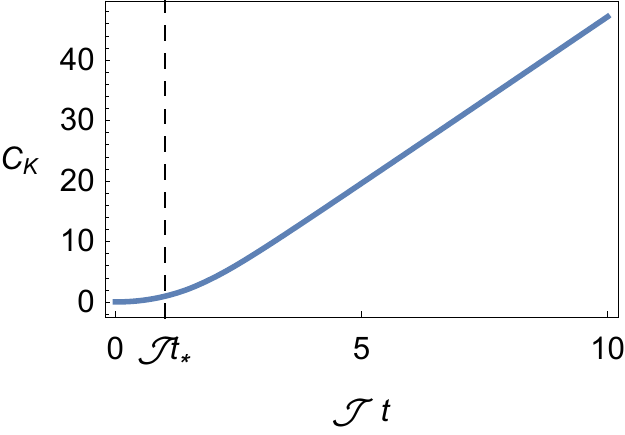}\label{fig:ck}} \quad \quad
\subfigure[]{
    \includegraphics[width=5.5cm]{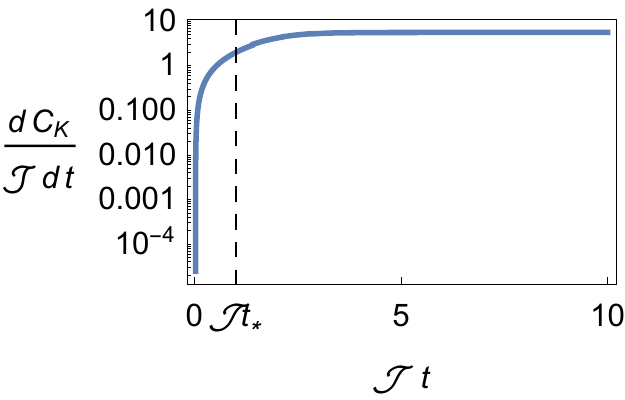}\label{fig:dck}}
	\caption{(a) The Lanczos coefficient of the SYK model. The curves from the bottom to the top correspond to $q=4$ and $N=20,22,24,26,28,30$, respectively.
	(b) The plateau value of the Lanczos coefficient at different $N$.
	$\bar b$ given by the plateau value of the Lanczos coefficient $n > N/q$.
	The dashed line is fitted by a linear function, showing that the platue value is proportional to $N$.
	(c) The K-complexity of the Heisenberg operator $\sqrt2 \psi_1 (t)$ in the SYK model. We use the parameters $q=4$ and $N=30$.
	(d) The time derivative of the K-complexity shown in (c). The scrambling time is denoted by $t_\ast$. Due to the small system size, the exponential growth is not obvious.}
\end{figure}

We now consider the example of SYK in detail. Our goal is to demonstrate the expectations~(\ref{eq:complexity}) explicitly. Once again, the SYK Hamiltonian is defined as
\bea
	H = \frac{i^{\frac{q}2}}{q!} \sum_{j_1,..., j_q} J_{j_1,...,j_q} \psi^{j_1}...\psi^{j_q}, \quad \overline{J_{j_1,...,j_q}^2} = \frac{(q-1)!J^2}{N^{q-1}} = \frac{2^{q-1}  (q-1)! \J^2}{q N^{q-1}}.
\eea
where the Majorana fermions satisfy $\psi^\dag_j = \psi_j$, and $\{\psi_{i}, \psi_{j} \} = \delta_{ij}$. Exponential growth of K-complexity at early times in the SYK model has been obtained analytically in the large-$q$ limit and numerically by solving the Schwinger-Dyson equation (see Appendix B in~\cite{Parker:2018a}). At large $q$, the operator wave function of a single Majorana fermion at time $t$ is
\bea
    \phi_n(t) = \sqrt{\frac{2}{n q}} \tanh^n \J t, \qquad n \ge  1,
\eea
leading to the exponential growth of K-complexity at early times,
\bea\label{SYK_C}
	C_K(t) = \sum_{n=1}^\infty n |\phi_n(t)|^2 =  \frac1q (\cosh 2\J t-1),
\eea
where the summation over the basis can be extended to infinity because we work at finite time with $N \rightarrow \infty $. The exponential growth exponent is $2\J$, consistent with the Lyapunov exponent at infinite temperature.

In getting (\ref{SYK_C}), the large-$N$ Wightman correlation function is used, but this only works for the K-complexity before the scrambling time. After the scrambling time, we expect a linear growth of K-complexity. To verify this conjecture, we choose $O = \sqrt 2 \psi_1$, and calculate the Lanczos coefficients numerically. For practical purposes, we can truncate the Krylov space at some $n_{\max} \gg 1$ and still capture the dynamics for a finite time related to $n_{\max}$. In Fig.~\ref{SYKLanczos}, we plot the Lanczos coefficients for different choices of $N$, with all showing plateau behavior in the regime $n> N/q$. Moreover, the value of the plateau is proportional to the system size $N$, which is shown in Fig.~\ref{fig:plateau}. (At exponentially large $n \sim e^{N}$, the Lanczos coefficient decreases and eventually vanishes because the dimension of Krylov basis is bounded by that of the operator Hilbert space~\cite{Rabinovici:2020operator}.) The K-complexity and its time derivative are calculated by solving the Schr\"odinger equation~(\ref{Schrodinger}) for $q=4$ and $N=30$, with the results shown in Fig.~\ref{fig:ck} and~\ref{fig:dck}. At late times, $t \gg t_* = \lambda_L^{-1} \log N/q$, the K-complexity grows linearly as expected. Due to the small number of qubits we simulate, $N/2=15$, the scrambling time is quite small, so the early time exponential regime is not manifest.

\begin{figure}
\subfigure[]{
	\includegraphics[height=4cm]{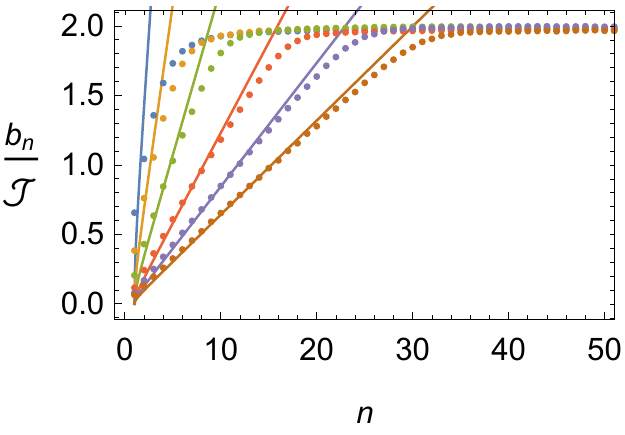}\label{fig:SYKLanczonsBeta}} \quad
\subfigure[]{
	\includegraphics[height=4cm]{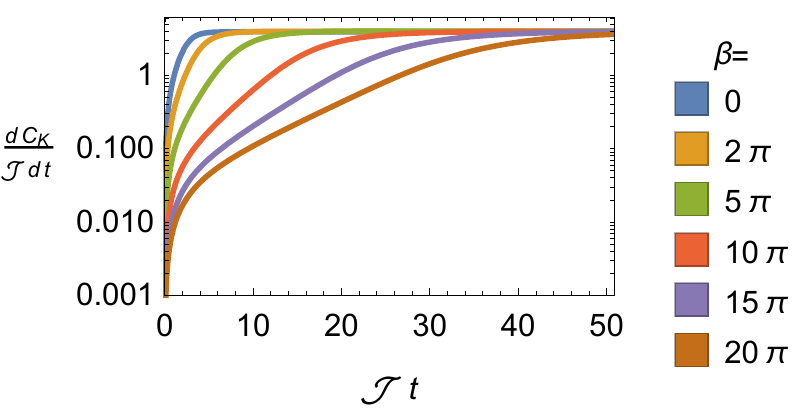}\label{fig:SYKComplexityBeta}} 
    \caption{(a) The Lanczos coefficient of the SYK model at different temperatures. The slope of the solid line is determined by $2\alpha(\beta)$. We use the parameters $N=30, q=4, \mathcal J=1/\sqrt2$.
    (b) The K-complexity growth of a single Majorana fermion at different temperature.}
\end{figure}

So far, we considered only the infinite temperature case. One way to generalize K-complexity to finite temperature is to consider the corresponding PETS $\ket{O_\beta} \equiv \ket{O_\beta(\theta=0)}$ at temperature $1/\beta$. We focus on $\theta = 0$, and the case of nonzero inserting angles $\theta \ne 0$ is considered in Appendix~\ref{appen:Ck_SYK}. This generalization is essentially equivalent to a change of definition of the inner product,
\bea
    \langle O_1 | O_2 \rangle_\beta = \Tr[e^{- \frac{\beta H}2 } O_1^\dag e^{- \frac{\beta H}2} O_2 ].
\eea
Ref.~\cite{Parker:2018a} obtained an analytic result at large $q$ prior to the scrambling time. We briefly review their analytic results at short times (see Appendix~\ref{appen:Ck_SYK}, and also the Appendix B in~\cite{Parker:2018a} for more details), and present a numerical evaluation valid at late times.

The moments are now related to the finite temperature Wightman correlation function,
\bea
    G(t) = \frac{\Tr[\rho^{1/2}O \rho^{1/2}O(t)]}{\Tr[\rho^{1/2}O \rho^{1/2}O]}, \qquad \rho = e^{-\beta H}, \qquad \mu_{2n} = i^{2n} \frac{d^{2n}}{dt^{2n}} G(t)|_{t=0},
\eea
Using the large-$q$ Wightman correlation function at finite temperature $\beta >0$~(Appendix~\ref{appen:Ck_SYK}), we get the generalized K-complexity at early times,
\bea\label{SYK_CBeta}
	C_K(t) = \sum_{n=1}^\infty n |\phi_n(t)|^2 =  \frac1q (\cosh 2\alpha t-1),
	\qquad
	\alpha = \J \cos \frac{\alpha\beta}2 \rightarrow \begin{cases} \J, \quad & \J \beta \ll 1 \\ \pi/\beta , \quad & \J \beta \gg 1 \end{cases}.
\eea
The exponential growth rate is given by $2\alpha$, which is equal to the Lyapunov exponent $\lambda_L = 2\alpha$ at large $q$~\cite{Maldacena:2016remarks}.

The analytic wavefunction from the large-$q$ Wightman correlation function also allows us to compute moments of $\C_K$ operator. To do that, we can introduce the generating function of K-complexity, i.e.,
\bea \label{eq:ck_generating}
    \avg{e^{\mu \C_K}} &=& 1+ \frac4q \log \sech \alpha t + \sum_{n=1}^\infty e^{\mu n} \frac2{n q} \tanh^{2n} \alpha t = (1+(1-e^\mu) \sinh^2 \alpha t )^{-2/q}.
\eea
The $n$-th moment is obtained by taking $n$-th derivative of the generating function, but we restrict ourselves to the average K-complexity. It would be interesting to explore the holographic duality of such an generating function in the future.

We also obtain the Lanczos coefficient at large $n \gg N$, as shown in Fig.~\ref{fig:SYKLanczonsBeta}, which shows a similar linear-to-plateau pattern. The slope of the linear function in the Lanczos coefficient gets smaller at lower temperature, because the dynamics is slower, reflecting the decrease of the Lyapunov exponent $\lambda_L = 2\alpha$ with temperature. The ETH estimate applied to the operator $O_\beta$ leads to $(O_\beta)_{ab} = A(E_a, E_a)  \delta_{ab} + A(E_a,E_b) 2^{-N/2} R_{ab}$ with $A(E_a,E_b)=A(0,0)F(E_a-E_b)e^{-\beta(E_a+E_b)/2}$. Similar to (\ref{muETH}), $\mu_{2n}\sim (N\mathcal E)^{2n}$ for $n\gg N$. As a result the plateau value of the Lanczos coefficient remains unaffected. The time derivative of K-complexity growth for various temperatures is plotted in Fig.~\ref{fig:SYKComplexityBeta}, where the slope of the late-time linear growth is independent from the temperature, and the early-time exponential growth region expands due to the decrease of the Lyapunov exponent at finite temperatures.

\section{Holographic complexity in the JT gravity} \label{sec:cv}

\subsection{Partially entangled thermal state in JT gravity}

We now carry out corresponding calculations in JT gravity using complexity-volume duality. First, recall the setup of JT gravity. For a holographic conformal field theory, the bulk representation of the PETS state is a half disk with an operator inserted on its boundary \cite{Maldacena:2001eternal,Goel:2018expanding}. We assume that the dual bulk theory is JT gravity~\cite{Maldacena:2016conformal} with a free matter field $\chi$ (the bulk field dual to operator $O$) coupling only to the metric,
\bea
I &=& I_{\rm bdy}[g, \phi] + I_M[g,\chi],\\
I_{\rm bdy}[g, \phi] &=& -\frac{\phi_0}{16\pi G_N}\kd{\int\sqrt g R+2\int_\partial K}-\frac1{16\pi G_N}\kd{\int d^2x \phi\sqrt{g}(R+2)
+2\int_\partial dx\sqrt{h}\phi_b K},\label{ActionJT}
\eea
where $G_N$ is the Newton's constant. $g$ and $h$ refer to the determinant of metric $g_{\mu\nu}$ and of induced metric. $R$, $K$, $\phi$, and $\phi_b$ denote the scalar curvature, the extrinsic curvature, the dilaton field and its value on the boundary, respectively. We require that the constant $\phi_0\gg \phi$. 
The boundary conditions are
\bea\label{JTBC}
    h = \frac1{\epsilon^2}d\tau^2, \qquad \phi_b = \frac{\phi_r}{\epsilon},
\eea
where $\phi_r$ is chosen to be a constant and $\tau$ is the imaginary time of the boundary theory.

The first term of (\ref{ActionJT}) is purely topological and gives the residual entropy $S_0=\phi_0/4G_N$. The second term in the bulk gives $R+2=0$ after the dilaton field $\phi$ is integrated out. So the metric in the bulk is localized to $EAdS_2$, $ds^2 = (d\tilde\tau^2 + dz^2)/z^2$ (in this section, we work in Euclidean signature). Thus, the remaining dynamics of the metric is on the boundary and is governed by the last term of~(\ref{ActionJT}). This can be effectively reduced to the reparametrization of the boundary time~\cite{Maldacena:2016conformal}
\bea\label{eq:reparameterization}
\tilde\tau=f(\tau),\quad z=\epsilon f'(\tau)+o[\epsilon^3],
\eea
which automatically satisfy the boundary condition of metric in (\ref{JTBC}). The time reparametrization field $f(\tau)$ is governed by the Schwarzian action \cite{Maldacena:2016conformal,Maldacena:2017diving},
\bea\label{eq:Schwarzian}
I_{\rm bdy}[f] 
&=&-Q\int d\tau\sqrt h (K-1)=-Q\kc{2\pi-  \int dx^2 \sqrt g \frac R2-\int d\tau\sqrt h } 
= -Q(2\pi+A -L)\\
&\approx& - \epsilon Q \int_0^\beta d\tau ~{\rm Sch}(f(\tau), \tau), \\
&&\text{where}~~
Q= \frac{\phi_b}{8 \pi G_N},\quad
L = \frac\beta\epsilon,\quad
{\rm Sch}(f(\tau), \tau) = - \frac12 \Big(\frac{f''}{f'}\Big)^2 +\Big( \frac{f''}{f'} \Big)'.
\eea
We have used the Gauss-Bonnet theorem and $R=-2$ in the first line. $A$ is the area enclosed by the boundary. $L$ is the length of the boundary.

We assume that the inserted operator $O$ is a single trace operator with scaling dimension $\Delta$ and the dual matter field $\chi$ vanishes in vacuum. We only consider the case of $\Delta>0$ which agrees with the situation in the SYK model. The dimensionless inner product of the PETS becomes \cite{Goel:2018expanding}
\bea
  &&\epsilon^{2\Delta} \frac{\avg{ O_\beta (\theta) | O_\beta (\theta)}} {\avg{\mathbb1_\beta|\mathbb1_\beta}}
  = \epsilon^{2\Delta} \avg{ O(\tau) O\kc{\tau'} }_\beta 
  = \epsilon^{2\Delta}\frac{\delta^2}{\delta\chi_r(\tau) \delta\chi_r(\tau')}\int Df~e^{-I_{\rm bdy}[f]- I_M^\eff[f,\chi]} \\
  &=&  \int Df \left(  \frac{\epsilon^2 f'(\tau) f'(\tau')}{[f(\tau)-f(\tau')]^2} \right)^\Delta e^{- I_{\rm bdy}[f]}
  \approx \int Df \kd{2\cosh D(X,X')}^{-\Delta} e^{-I_{bdy}[f]}, \\
&\tau &= \frac\beta4 \Big(1+ \frac{2\theta}\pi \Big),  \quad  \tau' = -\frac\beta4 \Big(1 - \frac{2\theta}\pi \Big),
\eea
where $X$($X'$) denotes the point on the boundary with the boundary time $\tau$($\tau'$), and $D(X,X')$ denotes the geodesic distance between the two points $X$ and $X'$. In getting the above equation, we have used the effective action of the matter field \cite{Maldacena:2016conformal} and the approximation of geodesic distance under the reparameterization (\ref{eq:reparameterization})
\bea
I_M^\eff[f,\chi]=-\int d\tilde\tau_1d\tilde\tau_2\frac{\tilde\chi_r(\tilde \tau_1)\tilde\chi_r(\tilde \tau_2)}{|\tilde\tau_1-\tilde\tau_2|^{2\Delta}}
,\quad \chi_r(\tau)=\kd{f'(\tau)}^{1-\Delta}\tilde\chi_r(f(\tau))
\\
\cosh D(X(\tau_1),X(\tau_2))
=\frac{(\tilde\tau_1-\tilde\tau_2)^2+z_1^2+z_2^2}{2z_1z_2}
=\frac{\kd{f(\tau_1)-f(\tau_2)}^2}{2\epsilon^2 f'(\tau_1)f'(\tau_2)}+ o[\epsilon^0].
\eea 
where we have normalized the prefactor of the correlation function in the Poincare coordinate, and $\chi_r(\tau)$ is the source of $O(\tau)$.

Combining the gravity part and the matter part, we obtain the effective action up to some constants
\bea\label{eq:JTEffAction}
    I = -Q(A-L) + \mu L_\mu,\quad
    \mu=\Delta, \quad  L_\mu \approx D(X,X').
\eea
Equivalently, the first part describes a massive particle with charge $Q$ moving in a hyperbolic space, where $A$ is the area enclosed by the world line and $L$ is the length of the world line~\cite{Maldacena:2017diving}. We call it the boundary particle because it locates the boundary. The second part describes a neutral particle, where $L_\mu$ and $\mu$ are the world line length and the mass of the inserted particle, respectively. We call it the inserted particle because it describes the operator insertion. So the degree of freedom are the trajectories of the boundary particles and the inserted particle, where the total length of the world line of the boundary particles $L$ is fixed to be $\beta/\epsilon$.

We also need to connect the world lines between the boundary particle, and the inserted particle according to the inserted position $\theta$. The dimensionless parameters in the problem are $\ke{Q,L,\mu,\theta}$, where $\ke{Q,L,\mu}$ are measured in units of the AdS radius.

\begin{figure}
\subfigure[]{	
	\includegraphics[height=5cm]{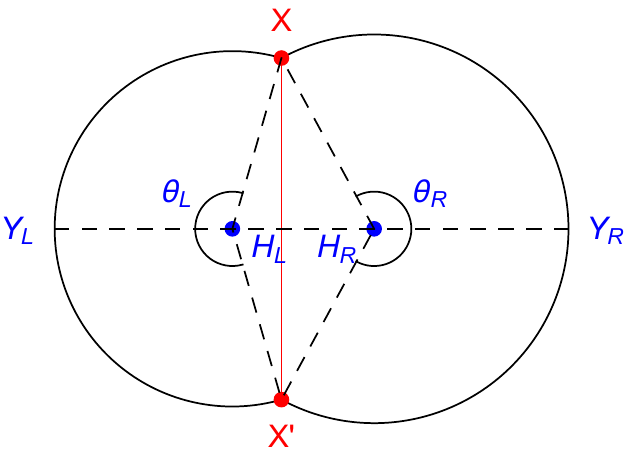}\label{fig:euclidean}} \qquad \qquad
\subfigure[]{
	\includegraphics[height=5cm]{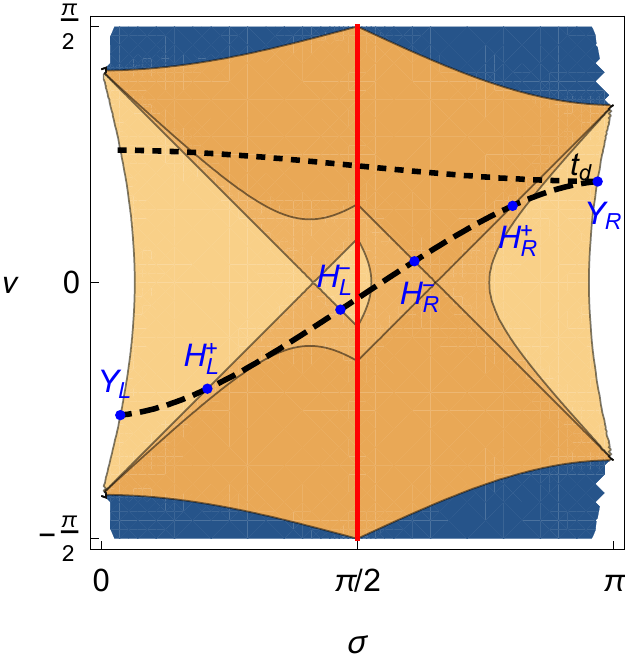}\label{fig:ads2}}
	\caption{(a) A schematic trajectory of boundary particles in hyperbolic space.
	The two segments and the vertical line represent the world lines of the two boundary particles and the inserted particle. The true horizon is point $H_R$.
	(b) The configuration of the dilaton field in the global coordinate of $AdS_2$ with the operator insertions. The parameters are $Q=100$, $L=30$, $\mu=20$ and $\theta=\pi/20$. We plot the geodesics of $C_V(-t_d,t_d)$ (dashed) and $C_V(t_d,t_d)$ (dotted). These geodesics intersect the left boundary, left-outer horizon, left-inner horizon, right-inner horizon, right-outer horizon, and right boundary at points $\ke{Y_L,\,H_L^+,\,H_L^-,\,H_R^-,\,H_R^+,\,Y_R}$.}
\end{figure}

Throughout we assume a low-energy limit and a classical limit $Q\gg L\gg 1$. To minimize the action, the classical solution tends to have a large $A$ and a small $L_\mu$. One can have an intuitive picture of the solution in Fig.~\ref{fig:euclidean}. We will discuss quantitatively below. 

To prepare the PETS, it is convenient to use an embeddeding space, $-(Y^{-1})^2+(Y^{0})^2+(Y^{1})^2 = -1$ with the metric $ds^2 = -(dY^{-1})^2+(dY^{0})^2+(dY^{1})^2$. Other coordinate systems are summarized in Appendix~\ref{appen:coord}. 

The action (\ref{eq:JTEffAction}) is invariant under the an overall SL(2) transformation on the left boundary particle, the right boundary particle, and the inserted particle. The SL(2) charges of the three particles are denoted as $\ke{Z_L^a,Z_R^a,Z_\mu^a}$ respectively.
Their trajectories $\ke{Y_L^a,Y_R^a,Y_\mu^a}$  are determined by their own SL(2) charges via \cite{Maldacena:2017diving,Lin:2019symmetries}
\bea\label{eq:trajectory}
	Z_R \cdot Y_R = -Q, \quad Z_L\cdot Y_L = Q, \quad Z_\mu \cdot Y_\mu=0.
\eea
Actually, the overall SL(2) transformation is a gauge redundancy due to the killing symmetries of the background $AdS_2$. So the total charges vanish
\bea \label{eq:conservation}
    Z_R^a+Z_L^a+Z_\mu^a = 0,
\eea
which controls the interactions between the three particles.
In the rest frame of the inserted particle, we make the following ansatz for the SL(2) charges,
\bea\label{eq:charge}
	Z_i^a = \frac{Q}{\cosh r_i} (s_i\cosh \rho_i, 0, -\sinh \rho_i),\quad
	Z_\mu^a = (0,0,-\mu),\quad s_{L,R}=-1,1,\quad i=L,R.
\eea
where $r_i$ and $\rho_i$ are constants determined by the equation of motion.
From (\ref{eq:trajectory}), the trajectories of the two boundary particles are
\bea
    Y_i(\varphi) = e^{-s_i \rho_i T_2} (\cosh r_i,~ \sinh r_i \sin \varphi,~ \sinh r_i \cos \varphi)^T,
    \quad T_2 = \left( \ba{cccc} 0 & 0 & 1 \\ 0 & 0 & 0 \\ 1 & 0 & 0 \ea \right),
\eea
which are parameterized by the angle $\varphi$ in the Rindler coordinates.

Schematic trajectories for these particles are shown in Fig.~\ref{fig:euclidean}: the two segments are world lines of the left and right boundary particles with centers at $H_L$ and $H_R$, respectively. The red line is the inserted particle. The segments are associated with angles $\theta_i$ which satisfy
\bea  \label{eq:rho}
    \tanh \rho_i = - \tanh r_i \cos \frac{\theta_i}2, \quad i = L, R,
\eea
because the three trajectories join at $X$ and $X'$. The length of the world lines of two boundary particles is fixed by
\bea \label{eq:theta}
	\theta_i \sinh r_i = \frac{L}2 \kc{ 1+s_i \frac{2\theta}\pi },
\eea
which also enforces the total length to be $L$. With Eqs.~(\ref{eq:conservation}), (\ref{eq:rho}) and (\ref{eq:theta}), we can solve for $\ke{\rho_i,r_i,\theta_i}$ in terms of the parameters $\ke{Q,L,\mu,\theta}$.

The dilaton field can be determined from its equation of motion. At point $Y$, one has
\bea
    \frac{\phi}{8\pi G_N} = \begin{cases}
    Z_L \cdot Y, & Y^1<0 \\
    -Z_R \cdot Y, & Y^1>0
    \end{cases}.
\eea
Figure~\ref{fig:ads2} shows the configuration of the dilaton field for the PETS in global coordinates for $AdS_2$. When the centers $H_L$ and $H_R$ are on the both sides of the trajectory of the inserted particle in Euclidean $AdS_2$, as shown in Fig.~\ref{fig:euclidean}, the dilaton field reaches extremal values at the two points. We call them horizons $H_L,\,H_R$ although only the one with the smaller value of dilaton is the true horizon \cite{Goel:2018expanding}. Each of the horizons extends along a light cone in Lorentzian $AdS_2$, as shown in Fig.~\ref{fig:ads2}. We call the left(right)-going light cone of the left horizon as the left-outer(inner) horizon, and the right(left)-going light cone of the right horizon as the right-outer(inner) horizon.

Using the conservation law~(\ref{eq:conservation}) and the constraints~(\ref{eq:rho},\ref{eq:theta}), the SL(2) charges of the boundary particles can be obtained numerically.
One can also can get analytic solutions perturbatively in $\frac{\mu L} Q \ll 1$,
\bea\label{solution}
	r_i = r - \frac{2+(\pi + 2\theta s_i)\tan\theta}{4\pi^2} \frac{\mu L}{Q},
\eea
where $\sinh r = \frac{L}{2\pi}$ is the unperturbed radius.

However, at $\theta=\frac\pi2$, the expansion~(\ref{solution}) breaks down. To get a meaningful result, one can introduce a regularization, i.e., $\theta = \frac\pi2 - \delta$, and expand first in $\delta \rightarrow 0$ and then in $\frac{\mu L^2}Q \ll 1$. The result is
\bea\label{solution2}
    r_R =r  , \qquad r_L = r -\frac1{4\pi^2} \frac{\mu L^2}Q.
\eea
In this case, the expansion parameter is $\frac{\mu L^2}Q$ which differs from the parameter $\frac{\mu L}{Q}$ at generic $\theta$. This reflects the non-commutativity of the two expansions in the $\delta = 0$ limit.

\subsection{Holographic complexity growth of the Heisenberg operator}

We define the holographic complexity $C_V(t) $ of the Heisenberg operator as the holographic complexity of the corresponding PETS. In two-dimensional spacetime, the CV conjecture~\cite{Stanford:2014complexity} states that the holographic complexity is proportional to the geodesic distance $D$ between the boundary points at times $t_L$ and $t_R$ in Lorentzian signature. It is approximated by \cite{Brown:2018JT}
\bea
	C_V(t_L,t_R) \approx \frac{\phi_0}{G_N} D\kc{ Y_L\kc{\pi-i u_L}, Y_R\kc{i u_R}},\quad
	u_i=\frac{2\pi t_i}{\beta_i},\quad
	\beta_i=\frac{\pi+2\theta s_i}{\theta_i}\beta,
\eea
where $\phi_0$ dominates the cross-section. The geodesic distance $D(Y_1,Y_2)$ between points $Y_1$ and $Y_2$ can be evaluated by the inner product in the embeddeding space,
$\cosh(D( Y_1, Y_2))=- Y_1 \cdot Y_2 $. Finally, letting $\tilde C_V=\frac{G_N}{\phi_0} C_V$, we have
\bea\label{eq:chD}
\cosh\tilde C_V(t_L,t_R)
&=&\cosh \left(\rho _L+\rho _R\right) \kc{ \cosh r_L \cosh r_R +\sinh r_L \sinh r_R \cosh u_L \cosh u_R } \nn\\
&& - \sinh \left(\rho _L+\rho _R\right) \kc{ \sinh r_L \cosh r_R  \cosh u_L + \cosh r_L \sinh r_R \cosh u_R } \nn \\
&& + \sinh r_L \sinh r_R \sinh u_L \sinh u_R.
\eea

Consider first the case with $-t_L=t_R=t$. The choice of opposing directions of time evolution corresponds to the Heisenberg evolution of operators, $U(t)^\dag O_\beta(\theta) U(t)$.
Without the inserted particle, $\mu=0$, the bulk is unperturbed and recovers the Rinder patch of $AdS_2$, $\rho_L + \rho_R  = 0$, $r_L = r_R = r$ and $-u_L = u_R = \frac{2\pi t}{\beta}$.
The complexity is $C_V[\mathbb{1_\beta}]=2\phi_0r/G_N$, which is independent of time because of the boost symmetry of the TFD state.

In the light operator limit, $\mu L \ll Q$, and considering $|\theta|\neq \pi/2$, the perturbed solution~(\ref{solution}) can be used to get the geodesic length,
\bea\label{eq:length}
	\cosh\tilde C_V(-t,t) &=& \begin{cases}
	    \frac{e^{2r}}2 \Big( 1 + \frac{\pi \sec \theta \cosh u - (2 + 2 \theta \tan \theta) }{2\pi^2}  \frac{\mu L}{Q} \Big), \quad & t_d \ll t \ll t_\ast \\
	     \frac12 (\frac{\sec \theta}{8\pi} \frac{\mu L}{Q}e^{r+u})^2  , \quad & t\gg t_* \end{cases},\quad
	     u=\frac{2\pi}{\beta}t
\eea
where $r \sim \log L$, and the dissipation time and the scrambling time are, respectively,
\bea \label{timescale}
t_d=\frac{\beta}{2\pi},\quad t_*=\frac{\beta}{2\pi}\log\frac{8\pi Q\cos\theta}{\mu L}.
\eea
At late times, we only need consider the leading time dependence in~(\ref{eq:chD}). The complexity grows exponentially at early times and linearly at late times,
\bea
	\tilde C_V(-t,t) \approx  \begin{cases}
		2r + \frac{ \sec \theta}{2\pi} \frac{\mu L}{Q} \cosh u, \quad & t_d \ll t \ll t_\ast \\
	    2\log \Big(\frac{ \sec \theta}{8\pi} \frac{\mu L}{Q} \Big) + 2r + 2u, \quad  \quad & t\gg t_*,
		\end{cases}
\eea
The early time exponential growth has Lyapunov exponent $\lambda_L = 2\pi/\beta$, and at late time the linear growth rate of $\tilde C_V$ is $\lambda_C= 4\pi/\beta$.

In the $\theta = \frac\pi2$ case, we use instead the perturbed solution~(\ref{solution2}), and expand the geodesic length in the limit $\delta \rightarrow 0$ and $\mu L^2 \ll Q$,
\bea
	\cosh\tilde C_V(-t,t) &=& \begin{cases}
	    \frac{e^{2r}}2 \Big( 1 + \frac{\cosh u -1}{4\pi^2} \frac{\mu L^2}Q \Big),  \quad & t_d \ll t \ll t_\ast \\ 
	    \frac12 \Big(\frac{1}{2\pi} \frac1{8 \pi}\frac{\mu L^2}{Q} e^{r+u} \Big)^2, \quad & t\gg t_*
	    \end{cases}, \quad  u=\frac{2\pi}{\beta}t,
\eea
leading to the following complexity dynamics,
\bea
    \tilde C_V(-t,t) \approx \begin{cases}  2r + \frac{1}{4\pi^2} \frac{\mu L^2}Q \cosh u, \quad & t_d \ll t \ll t_\ast \\
    2\log\Big( \frac1{2\pi}\frac{1}{ 8\pi} \frac{\mu L^2}Q \Big) + 2r  +
    2u,  \quad & t\gg t_*
    \end{cases}.
\eea
In particular, the qualitative behavior is not modified by the change in perturbation parameter and the Lyapunov exponent and late time exponent are still given by $\lambda_L = 2\pi/\beta$ and $\lambda_C= 4\pi/\beta$, respectively. It is also interesting to note that if one regularizes the infinity at $\frac\pi2$ by $\sec \frac{\pi}2 \rightarrow \frac{L}{2\pi}$, then (\ref{eq:length}) can include the case $\theta \rightarrow \frac\pi2$ as well.

\begin{figure}
\subfigure[]{\label{fig:cvilight}
	\includegraphics[height=4cm]{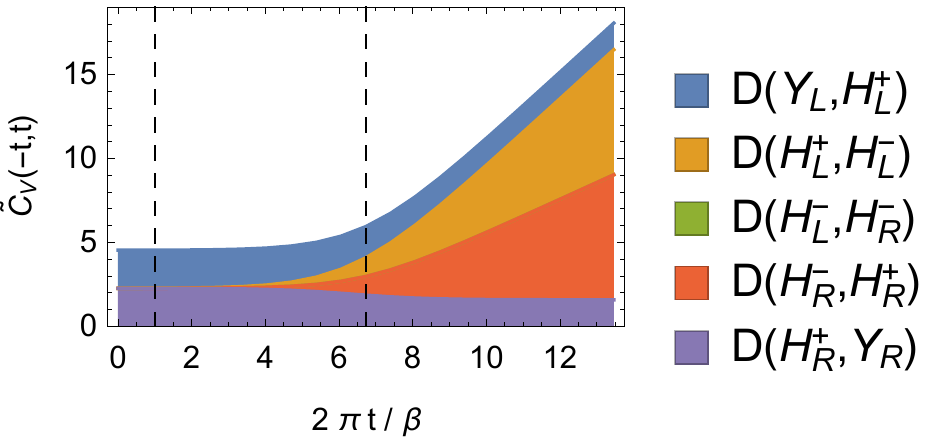}}
\subfigure[]{\label{fig:dcvlight}
	\includegraphics[height=4cm]{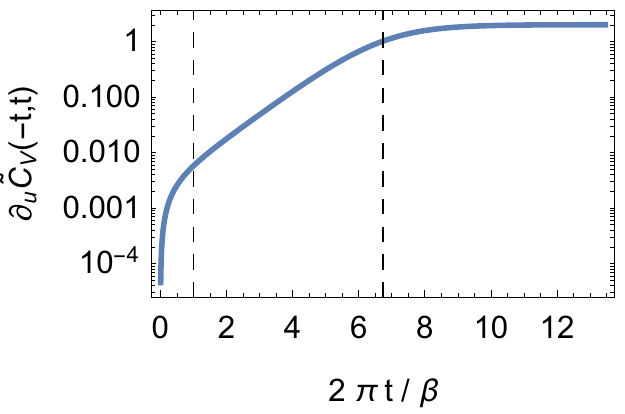}}
	\caption{ (a) The complexity growth of a light Heisenberg operator, where the contributions from the intervals between points $\ke{Y_L,\,H_L^+,\,H_L^-,\,H_R^-,\,H_R^+,\,Y_R}$ are shown, where the contribution from $D(H_L^-,H_R^-)$ is too small to be seen. (b) The time derivative of the complexity growth. The two dashed lines indicate the dissipation time $t_d$ and the scrambling time $t_\ast$ in~(\ref{timescale}). The parameters are $Q=100$, $L=30$, $\theta=0$, and $\mu=0.1$.
	\label{fig:cvlight}}
\end{figure}

These analytic results are also consistent with a numerical evaluation of the complexity as shown in Fig.~\ref{fig:cvlight}. The numerical solution indeed shows that the complexity grows exponentially at first and linearly after the scrambling time.

Now, when $H_L,\,H_R$ are on both sides of the trajectory of the inserted particle, the geodesic giving the complexity crosses the left/right-inner/outer horizons at points $\ke{H_L^+,\,H_L^-,\,H_R^-,\,H_R^+}$ from left to right, as shown in Fig.~\ref{fig:ads2}. Combining with the two points $\ke{Y_L,\,Y_R}$ on the boundaries, they divide the complexity geodesic into five intervals. We plot the contributions of these intervals to the complexity in Fig.~\ref{fig:cvlight} in the light operator limit. The growth of complexity is mainly due to the growth of the length of the geodesic distances inside the black hole interiors $D(H_L^+,H_L^-)$ and $D(H_R^-,H_R^+)$. Note in particular the similarity between Fig.~\ref{fig:cvlight} and the K-complexity in Figs.~\ref{fig:ck},~\ref{fig:dck}.

To compare with the SYK model, we identify $\epsilon = 1/\J$. Then the parameters of the JT gravity and the SYK model are related by
\bea\label{JTSYK}
Q= \alpha_S N,\quad L = \beta \J,\quad \mu = \frac1q,\quad \frac{\phi_0}{4G_N} = s N,
\eea
$\alpha_S$ and $s$ are some numerical constants. At the large $q$ limit, $\alpha_S=\frac1{4q^2}$ and $s=\frac12\log 2$ \cite{Maldacena:2016remarks}. Based on~(\ref{SYK_CBeta}), if the boundary length segments are subtracted from the holographic complexity, we get the following equality in the conformal limit,
\bea\label{eq:CVCK}
	\frac{\Delta C_V(t) }{L} =  \frac{\alpha_C}q (\cosh \frac{2\pi t}\beta-1) = \alpha_C C_K(t), \quad \alpha_C = \frac{2s}{\pi \alpha_S}, \quad t_d < t < t_\ast.
\eea
where we have set $\theta = 0$ for comparison.
(Actually we verify in Appendix~\ref{appen:Ck_SYK} that for a nonzero $\theta$, the equality still holds.) 
Note that the scrambling time for K-complexity,  $\frac{\beta}{2\pi} \log \frac{N}q \frac{\J}\alpha \approx t_\ast \big( 1+ O[(\log N)^{-1}] \big)  $, is approximately equal to the scrambling time of the holographic complexity in the large $N$ limit. It is also interesting to compare the rate of complexity growth for both K-complexity and holographic complexity. In the conformal limit, the following equation hold after the scrambling time,
\bea\label{eq:rate}
    \frac{d \log \Delta C_V(-t,t)}{d t} = \frac{d \log C_K(t)}{d t}, \quad t> t_d.
\eea
At this level, the K-complexity defined in~(\ref{eq:ck}) thus gives a microscopic counterpart of holographic complexity.

\begin{figure}
    \includegraphics[height=4cm]{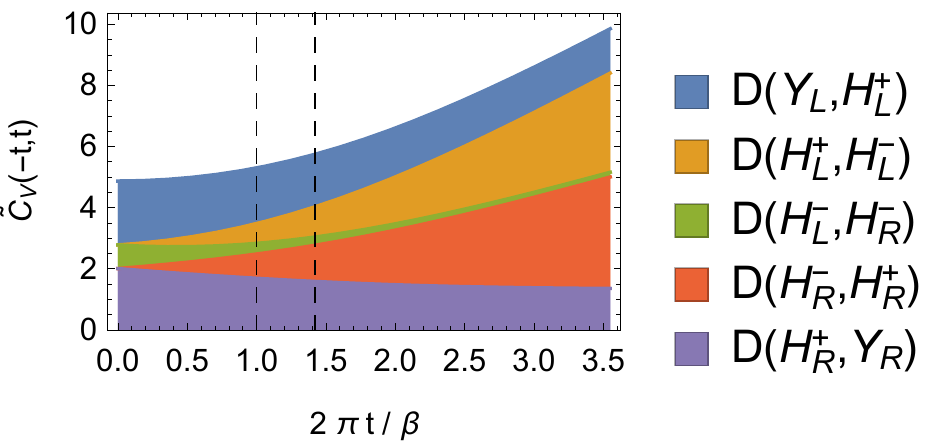}
    \caption{The complexity growth of a heavy Heisenberg operator. The parameters are $Q=100$, $L=30$, $\mu=20$ and $\theta=\pi/20$.}
    \label{fig:cvheavy}
\end{figure}

If two sides are evolved in the same time direction, then the complexity will grow linearly at first since the complexity of the time evolution operator $U$ will dominate over the simple operator $O$. The geodesic length and complexity are
\bea
	\cosh D(t,t) &=& \frac{e^{2r}}2 \cosh^2 u \Big(1 + \frac{\pi \sec \theta \sech u - (2 + 2 \theta \tan \theta) }{2\pi^2}  \frac{\mu L}{Q} \Big), \\
	 \tilde C_V (t,t) &\approx& 2r + 2u, \quad t \gg t_d, \quad u = \frac{2\pi t}\beta.
\eea
At late times, this growth is equivalent to the late-time linear growth of the complexity in simple Heisenberg operators. This is because the Heisenberg operator becomes complicated after the scrambling time, and the Heisenberg evolutions at two sides become mostly uncorrelated.

Another interesting limit is the heavy operator limit. When $\mu=2Q$, the two segments in Fig.~\ref{fig:euclidean} become two tangent thermal circles now with inverse temperatures $\beta_i=\frac{\pi+2\theta s_i}{2\pi}\beta$, respectively. The operator $O$ is already complicated initially, and there is not a large separation between the scrambling time and the dissipation time, as shown in Fig.~\ref{fig:cvheavy}. In this limit, the complexity approximately decomposes into the complexities of two wormholes separately,
\bea \label{CVheavy}
\tilde C_V(t_L,t_R)
&\approx& 2\log\kc{\frac{\beta_L}{\pi\epsilon}\cosh\frac{\pi t_L}{\beta_L}} + 2\log\kc{\frac{\beta_R}{\pi\epsilon}\cosh\frac{\pi t_R}{\beta_R}} \\
&=& \tilde C_V[e^{-\kc{\frac12\beta_L+it_L} H}] + \tilde C_V[e^{-\kc{\frac12\beta_R+it_R} H}]. \nn
\eea
This form exhibits quadratic growth when $t\ll t_d$ and linearly growth when $t\gg t_d$.
The decomposition reflects the fact that the heavy operator $O$ effectively cuts the wormhole into two shorter wormholes while creating a large interior, similar to a Python's lunch geometry~\cite{Brown:2019python}, as shown in Fig.~\ref{fig:ads2}. The geodesic length at two wormholes grows with time independently with their own inverse temperature $\beta_i$.

\section{The size of the partially entangled thermal state} \label{sec:size}

\subsection{The size from the SYK model}

In this section we obtain some results for operator size and compare them to the preceding complexity results. The maximally entanglement state $\ket 0$ defined in the doubled SYK Hilbert space satisfies $c^j\ket 0=0,\,\forall j$ where $c^j=(\psi_L^j+i\psi_R^j)/\sqrt2$. The size of an operator $\psi$ in the SYK model can be defined as $n[\psi]=\bra{\psi}\hat n\ket{\psi}/\avg{\psi|\psi}$, where the size operator is $\hat n=\sum_{j=1}^N (c^j)^\dag c^j= N/2+i\sum_j\psi^j_L\psi^j_R$ \cite{Qi:2018quantum}. At large $q$ and early time, the growth of size of operator $O_\beta(\theta+iu)=\sqrt2\psi^1_\beta(\theta+iu)$ is characterized by
\bea\label{SizeSYK}
\Delta\tilde n(t)=\frac{n[O_\beta(\theta+iu)]-n[\mathbb 1_\beta]}{\delta_\beta}
=\sec \left(\frac{\pi  v}{2}\right) \sec (\theta  v) \cosh (\frac{2\pi v}{\beta}t) -\frac{2 \tan \left(\frac{\pi  v}{2}\right) (\theta  v \tan (\theta  v)+1)}{\pi  v+2 \cot \left(\frac{\pi  v}{2}\right)},
\eea
where $v=\alpha\beta/\pi$, $u=2\pi t/\beta$, and the normalization factor $\delta_\beta=2G(\frac\beta2)$ is determined by $\Delta\tilde n(0)=1$ when $\theta=\frac\pi2$. At late time, the exponential growth will slow and eventually vanish as the size approaches $\Delta\tilde n(\infty)=N/2$. More generally, one can obtain the generating function of size operator at arbitrary inserting angle $\theta$, and it turns out the generating function of size operator agrees with that of K-complexity~\ref{appen:generating}.

Notice that, in the large $q$ limit, the scrambling time obtained from both $C_K$ and $C_V$ is $t_*\sim\lambda_L^{-1}\ln(N/q)$ rather than $\lambda_L^{-1}\ln N$. So $\Delta\tilde n(t_*)=N/q$ rather than its saturation value $N/2$, which implies that $\Delta\tilde n(t_*)$ deviates from exponential growth before saturation. We will discuss this point in Appendix \ref{SectionScramblingTime}.

\subsection{The size derived from JT gravity}

The size is related to the symmetries of $AdS_2$ in JT gravity. The SL(2) generators ${\tilde B,\tilde E,\tilde P}$ in the dual $AdS_2$ are related to the operators in the two sites SYK model \cite{Lin:2019symmetries}
\bea
\hat B&=&\frac\beta{2\pi}\kc{H_R-H_L},\\
\hat E&=&\frac\beta{2\pi}\kd{H_R+H_L+\tilde\mu\hat n - \bra{\mathbb 1_\beta} (H_R+H_L+\tilde\mu\hat n) \ket{\mathbb 1_\beta}}, \\
\hat P&=&-i[\hat B,\hat E],\\
&& \text{where}\quad \frac{\tilde\mu}{\mathcal J}=\frac{2\alpha_S}{\Delta\delta_\beta}\kc{\frac{2\pi}{\beta\mathcal J}}^2.
\eea
For conciseness, we consider that the states are normalized. The normalized change of size can be written as
\bea
\Delta\tilde n(t)
&=& \frac1{\delta_\beta} \kd{\bra{ O_\beta(\theta+iu)}\hat n\ket{O_\beta(\theta+iu)} - \bra{\mathbb 1_\beta}\hat n\ket{\mathbb 1_\beta}} \nn \\
&=& \frac{1}{\tilde\mu\delta_\beta}\frac{2\pi}{\beta} \kd{\bra{ O_\beta(\theta+iu)}(\hat E-\hat B-\frac{\beta}{\pi}H_L)\ket{O_\beta(\theta+iu)} + \frac{\beta}{\pi}\bra{\mathbb 1_\beta}H_L\ket{\mathbb 1_\beta}}  \\
&=& \frac{\Delta\mathcal J}{2\alpha_S}\frac{\beta}{2\pi} \kd{\bra{ O_\beta(\theta+iu)}(\hat E-\hat B)\ket{O_\beta(\theta+iu)} + \partial_\varphi\ln \left.G\kc{\frac\beta2\kc{1-\frac{2\theta}\pi};\beta\kc{1+\frac\varphi\pi}}\right|_{\varphi=0}} \nn,
\eea
where $G(\tau;\beta)=\Tr[e^{-\beta H}O^\dag(\tau)O]/\Tr[e^{-\beta H}]$.
Its time derivative is proportional to the momentum
\bea
\partial_t \Delta\tilde n(t)=\frac{\Delta\mathcal J}{2\alpha_S}\bra{O_\beta(\theta+iu)}\hat P \ket{O_\beta(\theta+iu)}.
\eea
According \cite{Lin:2019symmetries}, when we insert operators to the thermal circle at $\frac\pi2-\theta+iu$ and $-\frac\pi2+\theta+iu$, the generators in the semiclassical limit are
\bea
\frac{\bra{O_\beta(\theta+iu)}\kc{\hat B,\hat P,\hat E}\ket{O_\beta(\theta+iu)}}{\avg{O_\beta(\theta+iu)|O_\beta(\theta+iu)}} 
=\frac{\Delta}{\cos\theta}\kc{\sin\theta,\sinh u,\cosh u} 
\eea
At large $q$ limit, we obtain
\bea
\Delta\tilde n(t)= \frac{\beta\mathcal J}{\pi^2} (-2-\pi\tan\theta +\pi\sec\theta\cosh \frac{2\pi t}{\beta}),
\eea
which is equal to (\ref{SizeSYK}) from the SYK model at $\beta\mathcal J\gg1$ limit.

\subsection{Relation between the operator size and the complexity}

We first discuss the relation between K-complexity and the size operator. At infinite temperature, the time derivative of K-complexity is proportional to the size at early times $t_d \ll t \ll t_\ast$. Relating (\ref{SYK_CBeta}) and (\ref{SizeSYK}), we find that
\bea
    \frac1{2\J} \frac{d C_K(t)}{dt} \approx \frac{1}q \Delta \tilde n(t), \quad \beta \rightarrow 0.
\eea 
Actually, the generating functions of K-complexity and size operator agree, as shown in Appendix~\ref{appen:generating}. For finite temperature and $\theta=0$, the relation will be modified by a temperature dependent factor,
\bea
    \frac1{2\J} \frac{d C_K(t)}{dt} \approx \frac{1}q \kc{ \cos^2 \frac{\pi v}2 }  \Delta \tilde n(t).
\eea
This relation may be extended to late times after the scrambling time.
The linear growth of K-complexity at late times is proportional to the system size, i.e., $\frac1{2\J}\frac{d C_K(t)}{dt} \propto \frac{N}2 = \Delta n(t)$ for $t \gg t_\ast$.

Now we consider the relation between the holographic complexity and the size operator.
The size of an operators is linearly related to its out of time order correlator (OTOC) with Majorana fermions $\psi_i$ \cite{Qi:2018quantum}. From the geometric interpretation of effective theory \cite{Mertens:2017solving,Goel:2018expanding}, we find the following relation between the size in the SYK model and the complexity in JT gravity at the limit $q\gg1$, $N\gg\beta\mathcal J\gg1$ and the early time,
\bea
\frac{\pi ^2}{\beta\mathcal J}\frac{\Delta\tilde n[\psi^1_\beta(\theta+iu)]}{N/2}
=-2-2 \theta\tan\theta +\pi\sec\theta \cosh t
=\frac{2\pi^2 Q}{\mu L} \frac{G_N}{\phi_0}(C_V[\psi^1_\beta(\theta+iu)]-C_V[\mathbb 1_\beta])
\eea
which is valid under the dictionary (\ref{JTSYK}). Combining it with the Epidemic relation
$\frac d{dt}\Delta\tilde n = \lambda_L \Delta\tilde n$ at the Lyapunov regime \cite{Qi:2018quantum}, we find
\bea
\frac1{T S_0}\frac{d C_V}{dt} = \frac{\Delta\tilde n}{N/q},
\eea
where entropy $S_0=\frac{\phi_0}{4G_N}= sN$ and temperature $T=1/\beta$.

\section{Conclusion and outlook}

We calculated the complexity of a Heisenberg operator in both the SYK model and JT gravity. In the SYK model, we used the notion of K-complexity defined through the Krylov basis. In the JT gravity model, we used the CV conjecture to define the complexity. The simplicity of JT gravity allowed us to treat the problem of gravitational back-reaction by mapping it to motions of particles in a rigid hyperbolic space. We found that both complexities show an exponential-to-linear growth behavior. In particular, the two notions of complexity actually match up to a constant before the scrambling time. After the scrambling time, although the characteristic energy scales for the two complexities are different, they both show a linear growth with a slope proportional to system size. We also verified the relation between the complexity growth and the operator size before the scrambling time. The complexity can be used to capture the quantum dynamics at both short and long times.

It is worth noting the temperature dependence of the holographic complexity. For a generic insertion $\theta \ne \pm \pi/2$, the complexity is inversely proportional to the temperature, $C_V \propto (\beta/\epsilon) e^{2\pi t/\beta} $ before the scrambling time, and it is proportional to the temperature in the linear growth regime $C_V \propto ({\phi_0}/{G_N}) {4\pi t}/{\beta}$ after the scrambling time. The former is due to the fact that the inserted operator is perturbing the wormhole with temperature $1/\beta$, while the latter is due to the relation between the Lorentzian time and the Rindler time.

By contrast, the role of temperature is less clear in the context of computational complexity, so it is interesting to attempt to introduce temperature into the definition of computational complexity. For example, the temperature dependence of the holographic complexity means a simple identification of the circuit time and the Lorenzian time is not enough to set the complexities equal up to an overall constant. In our study, we generalized the K-complexity to finite temperature by considering the PETS at temperature $1/\beta$ as the reference state. This is essentially the same as a temperature-dependent inner product~\cite{Parker:2018a}. However, as we showed based on~(\ref{eq:CVCK}), such a generalization does not give a completely consistent identification between the two notions of complexity after the scrambling time. By adjusting the overall normalization of one or the other, one could match the early or late time growth but not both. Hence, it is interesting to consider further refinements that might produce even more harmony between the two notions of complexity. Nevertheless, we emphasize that at the level of the rate of complexity growth, the identification between the notions works perfectly well at early and late times as in~(\ref{eq:rate}). This suggests that as far as the temperature dependence is concerned, the rate of complexity growth ratio and the related time scales have simpler holographic interpretation than the absolute value of complexity itself.

A possible reason for the above mismatch is the difference between the reference states of K-complexity and holographic complexity. The reference state of K-complexity is the PETS state $\ket{O_\beta}$; the reference state of holographic complexity is the maximally entangle state $\ket0$~\cite{Stanford:2014complexity}. It is worth developing an algorithm of complexity which is independent from the choice of initial operator $O$. It will also benefit the generalization on the complexity of multi operators, which is related to multi shock waves geometries~\cite{Stanford:2014complexity}.

The CV proposal used in this paper only depends on the geometry and the dilaton. It is an open question that whether matter fields should have a direct contribution to the holographic complexity besides their indirect contribution via back-reaction on the metric. The answer to this question may be crucial for the complexity of heavy operators, such as~(\ref{CVheavy}). From the perspective of complexity-action (CA) conjecture \cite{Carmi:2016comments,Carmi:2017time,Brown:2015equalaction,Brown:2015action,Brown:2018JT,Cai:2017sjv,Cai:2016xho,An:2018Gauss-Bonnet}, the action of the matter field along the trajectory of the inserted particle can directly contribute. We hope to explore this problem in the future.

It is also interesting to consider higher dimensional generalization of PETS and its gravity dual. For instance, in three dimensions, we may insert an end of world brane behind the horizon of an eternal black hole, corresponding to the geometry worked out in~\cite{Balasubramanian:2020secret} in the context of an evaporating black hole. The holographic complexity in this case is then proportional to the volume of a two-dimensional maximal surface connecting the two boundaries. From the viewpoint of evaporating black holes, the geometry of the PETS in our study is effectively dual to a two-dimensional version of the entangled system consisted of the black hole and the auxiliary radiation~\cite{Balasubramanian:2020secret,Penington:2019replica}. And the holographic complexity calculated here is the so-called unrestricted complexity for decoding the radiation~\cite{Brown:2019python}.

\section*{Acknowledgement}
We thank Leonard Susskind, Anatoly Dymarsky, Matteo Carrega and Yu-Sen An for helpful discussions. S. K. J. and B. S. are supported by the Simons Foundation via the It From Qubit Collaboration. Z. Y. X. is supported in part by the Natural Science Foundation of China under Grants No.~11875053 and No.~12075298 and by the National Postdoctoral Program for Innovative Talents BX20180318, funded by China Postdoctoral Science Foundation.

\appendix

\section{K-complexity in the SYK model at early times} \label{appen:Ck_SYK}

We will summarize the Wightman correlation function and the K-complexity at early times (e.g. see Appendix B in~\cite{Parker:2018a} for more details).
Using the large-$q$ approximation, the imaginary time correlation function at temperature $\beta$ is given by
\bea
	2 \langle \mathcal{T}_\tau \psi(\tau)  \psi(0) \rangle = 1+ \frac2q \log \frac{\alpha}{\J |\cos\alpha (\tau - \beta/2)|}, \quad \tau>0, \quad \alpha = \J \cos \frac{\alpha\beta}2.
\eea
So, the Wightman correlation function is
\bea
    G(t) = 1+ \frac2q \log \frac{1}{ \cosh\alpha t}.
\eea
where we have properly normalize it by $G(0) = 1$. And accordingly the wavefunction of a simple Majorana fermion is~\cite{Parker:2018a}
\bea
	\phi_n(t) = \begin{cases} 1+\frac2q\log\frac{1}{\cosh(\alpha t)},& n=0 \\ \sqrt{\frac{2}{nq}} \tanh^n \alpha t, \quad & n \ge 1. 
	\end{cases}
\eea
This leads to the exponential growth of K-complexity at early time,
\bea
	C_K(t) = \sum_{n=1}^\infty n |\phi_n(t)|^2 =  \frac1q (\cosh 2\alpha t-1), \quad \alpha = \J \cos \frac{\alpha\beta}2.
\eea

We also generalize the analytical result of the K-complexity in the SYK model to non-zero inserting angles $\theta \ne 0$. For a nonvanishing $\theta \ne 0$, the diagonal component of the Liouvillian superoperator in Krylov basis is not zero,
\bea
    L_{mn} =  \delta_{n,m-1} b_{n+1} + \delta_{n,m+1} b_n + a_n \delta_{n,m}, \quad a_n = \langle [H,O_n] | O_n \rangle.
\eea
This change will lead to a slightly different Schr\"odinger equation for the wave function $O(t) = \sum_n i^n \phi_n(t) O_n$,
\bea
    \partial_t \phi_n = b_n \phi_{n-1} - b_{n+1} \phi_{n+1} + i a_n \phi_n.
\eea
The Lanczos coefficient can be mapped to a Toda chain flow and solved by a coupled differential equation~\cite{Dymarsky:2019quantum}. A general solution is obtained in~\cite{Dymarsky:2019quantum}, and after comparing with the correlation function from the SYK model, these coefficients read
\bea
a_n=\begin{cases}
2\alpha  \tan (v \theta)/q,& n=0\\
2 \alpha \tan (v \theta) n,& n\geq1
\end{cases}, \quad
b_n=\begin{cases}
 \alpha  \sec (v \theta) \sqrt{2/q}, & n=1\\
\alpha  \sec (v\theta) \sqrt{n (n-1)} , & n\geq2
\end{cases},\quad 
v=\frac{\alpha\beta}{\pi}.
\eea
It is directly to verify that the following wave function is the solution to the above Schr\"odinger equation,
\bea 
\phi_n(t)=\begin{cases}
1+\frac2q\log\frac{\cos(v\theta)}{\cos(i\alpha t+v\theta)},& n=0,\\
\sqrt{\frac2{qn}}\kc{\frac{\tanh(\alpha t)}{\cos(v\theta)-i\sin(v\theta)\tanh(\alpha t)} }^n, &n\geq1.
\end{cases}
\eea

It is suggective to include an angle dependent complexity to the definition of K-complexity for the SYK model, namely, 
\bea
    C_K = \sum_n \cos(\frac{\theta}\pi \alpha\beta) n |\phi_n|^2.
\eea
With this modification, the K-complexity of the SYK model for nonzero $\theta$ is
\bea
    C_K = \frac1q \sec(\frac{\theta}\pi \alpha\beta)(\cosh 2\alpha t-1).
\eea
At the conformal limit $\alpha \rightarrow \pi/\beta$, the K-complexity becomes
\bea
    C_K \approx \frac1q (\cosh \frac{2\pi t}\beta -1),
\eea
which agrees exactly with the computation from CV conjecture in JT gravity.

\section{Summary of coordinate systems}\label{appen:coord}
We summarize various coordinate systems used in the paper.
We start with embedding coordinate.
$AdS_2$ space can be embedded to
\bea
-Y_{-1}^2 - Y_0^2 + Y_1^2 = -1, \qquad ds^2 = -d Y_{-1}^2 - dY_0^2 + dY_1^2.
\eea
The global coordinate which we use to plot the perturbed $AdS_2$ spacetime is given by
\bea
	Y^{-1} = \frac{\cos \nu}{\sin \sigma}, \quad Y^0 = \frac{\sin \nu}{\sin \sigma}, \quad  Y^1 = \cot \sigma, \qquad ds^2 = \frac{- d \nu^2 + d \sigma^2}{\sin^2 \sigma}.
\eea
The Lorentzian coordinate system is related to the embedded coordinate by
\bea
	Y^{-1} = \frac{z}2 \Big[ 1 + \frac1{z^2} (1-\tilde t^2)\Big], \quad Y^0 = \frac{\tilde t}z , \quad Y^1 = \frac{z}2 \Big[ 1- \frac1{z^2} (1+ \tilde t^2)\Big], \qquad ds^2 = \frac{-d\tilde t^2 + dz^2}{z^2}.
\eea
Furthermore, a possible Rindler coordinate is
\bea
    Y^{-1} = \cosh r, \quad Y^0 = \sinh r \sinh \varphi, \quad  Y^{1} = \sinh r \cosh \varphi, \qquad ds^2 = d r^2 - \sinh^2 \rho d\varphi^2.
\eea

\section{Generating function for size at generic $\theta$} \label{appen:generating}

The generating function of size can be obtained in the large-$q$ limit. A simple generalization of \cite{Qi:2018quantum} gives the generating function of size operator at generic angle $\theta$,

\bea
&&
\frac{\bra{O_\beta(\varphi)}e^{\mu\hat n/\delta_\beta}\ket{O_\beta(\varphi)}}{\bra{\mathbb 1_\beta} e^{\mu\hat n/\delta_\beta} \ket{\mathbb 1_\beta} }
=e^{\mu} \kc{\frac{2\sin ^2\gamma _{\mu }}{-\cos \left(2 \gamma _{\mu }+ (\pi/2 - \theta ) v _{\mu }\right) +e^{q\mu} \cos \left((\pi/2 -\theta ) v _{\mu }\right)+\left(1-e^{q \mu}\right) \cosh \left(u v _{\mu }\right)}}^{2/q}, \nn \\
&&\pi v_\mu = \beta \mathcal J \sin\gamma_\mu, \quad 
e^{\mu q}\sin \left(\frac{\pi v_\mu}{2}\right)= \sin \left(2 \gamma_\mu +\frac{\pi v_\mu}{2}\right), \quad
\delta_\beta = \kc{\alpha/\J}^{2/q},\quad 
\alpha= \J\cos \alpha\beta /2,
\quad \varphi=\theta+iu. \nn\\
\eea
At infinite temperature $\beta \rightarrow 0$, the generating function reduces to
\bea
    \avg{ e^{\mu \hat n}} &=& \frac{e^{\mu}}{\Big[ 1  + (1 - e^{q \mu}) \sinh^2 \J t \Big]^{2/q}},
\eea
which agrees exactly with that of K-complexity in~(\ref{eq:ck_generating}) if one renormalizes the size by a factor of $q$ since by each step the Liouvillian the size increases a constant amount $q$. Note that the generating function works prior to the scrambling time since we implement the large-$q$ approximation.

\section{About the scrambling time}\label{SectionScramblingTime}

In this paper, the scrambling time is defined as the time of the crossover between the exponential growth and the linearly growth of complexities. We will show that this scrambling time also appears in the growth of the size, {\it i.e.} the decay of the OTOC, which should slow down before saturation.

The Schwarizan theory (\ref{eq:Schwarzian}) is able to capture the decay of OTOCs at both early time and late time.
In Ref.~\cite{Mertens:2017solving}, the OTOC $\avg{B_{l_2}(\tilde t_1)A_{l_1}(\tilde t_2)B_{l_2}(t_1)A_{l_1}(t_2)}$ corresponds to the gravitational scattering between the outgoing matter $A$ and the infalling matter $B$ near the horizon of the black hole with initial mass $m=\pi/\beta$. Semi-classically, assuming that the change in the mass of the black hole due to the matter is much smaller than $m$, and considering the small scaling dimensions $l_1,l_2\approx0$, one find that the time shift of the the outgoing matter $A$ is
\bea
\tilde t_2-t_2\approx \frac1{\lambda_L} \ln\kc{1+4\alpha C e^{\lambda_L(t_2-t_1-t_R)}},
\eea
where $C=\frac{\phi_r}{8\pi G_N}=\frac{Q}{\epsilon}=\frac{\alpha_S N}{\mathcal J}$, $m+\alpha$ is the mass of the black hole before the matter $A$ goes out, and $t_R=(2m)^{-1}\ln(4mC)$. The exponential time shift slows down when $1\sim 4\alpha C e^{\lambda_L(t_2-t_1-t_R)}$, namely at the scrambling time
\bea
t_*\sim \frac1{\lambda_L} \ln \frac{m}{\alpha}\sim \frac1{\lambda_L} \ln \frac Nq,
\eea
where we find the correspondence of $\alpha$ in the SYK model by matching the energies of PETS on both sides, {\it i.e.} $4Cm\alpha=\bra{\psi^1_\beta}H\ket{\psi^1_\beta}-\bra{\mathbb 1_\beta}H\ket{\mathbb 1_\beta}=2\mathcal J/q$. The $\ln(N/q)$ dependence in the scrambling time read from the OTOC agrees with the result of the complexities.

\bibliography{refs}

\end{document}